\documentclass[amsmath,showpacs,aps,prb,twocolumn,floatfix]{revtex4-1}

\usepackage{bbold}
\usepackage{graphics}
\usepackage{graphicx}
\usepackage{amsthm}
\usepackage{amsmath}
\usepackage{wasysym}
\usepackage{amssymb}
\usepackage{enumerate}
\usepackage{xfrac}
\usepackage{color}
\usepackage[dvips]{epsfig}

\usepackage[usenames,dvipsnames]{xcolor}
\definecolor{DarkBlue}{rgb}{0,0.08,0.45}

\begin{document}

\title{Entanglement Entropy and Spectra of the One-dimensional Kugel-Khomskii Model}

\author{Rex Lundgren}%
\email{rexlund@physics.utexas.edu}%
\author{Victor Chua}%
\author{Gregory A. Fiete}
\affiliation{Department of Physics, The University of Texas at Austin, Austin, TX 78712, USA}

\date{\today}

\begin{abstract}
We study the quantum entanglement of the spin and orbital degrees of freedom in the one-dimensional Kugel-Khomskii model, which includes both gapless and gapped phases, using analytical techniques and exact diagonalization with up to 16 sites. We compute the entanglement entropy, and the entanglement spectra using a variety of partitions or ``cuts" of the Hilbert space, including two distinct real-space cuts and a momentum-space cut.  Our results show the Kugel-Khomski model possesses a number of new features not previously encountered in studies of the entanglement spectra. Notably, we find robust gaps in the entanglement spectra for both gapped and gapless phases with the orbital partition, and show these are not connected to each other. We observe the counting of the low-lying entanglement eigenvalues shows that the ``virtual edge" picture which equates the low-energy Hamiltonian of a virtual edge, here one gapless leg of a two-leg ladder, to the ``low-energy" entanglement Hamiltonian breaks down for this model, even though the equivalence has been shown to hold for similar cut in a large class of closely related models.  In addition, we show that a momentum space cut in the gapless phase leads to qualitative differences in the entanglement spectrum when compared with the same cut in the gapless spin-1/2 Heisenberg spin chain. We emphasize the new information content in the entanglement spectra compared to the entanglement entropy, and using quantum entanglement present a refined phase diagram of the model. Using analytical arguments, exploiting various symmetries of the model, and applying arguments of adiabatic continuity from two exactly solvable points of the model, we are also able to prove several results regarding the structure of the low-lying entanglement eigenvalues.
\end{abstract}



\maketitle

\section{Introduction}
Over the last few years, quantum information theory has come to play a major role in deepening our understanding of strongly correlated quantum many-body systems.\cite{Eisert:rmp10,Amico:rmp08}  The focus of these studies has largely been on the entanglement {\em entropy} which is a single number obtained from the reduced density matrix of a sub-system of the larger quantum system of interest.\cite{Cheong:prb04,Vidal:prl03,Refael:prl04,Kitaev:prl06,Levin:prl06,Flammia:prl09}  However, the entanglement spectrum ({\it i.e.} the full set of eigenvalues of the reduced density matrix)\cite{Li:prl08} has emerged as a powerful tool to study strongly correlated quantum systems due to its ability to reveal subtle topological effects, and its direct connection in some cases to the physical boundary excitations.\cite{Chandran:prb11,Qi:prl12,Cirac:prb11,Peschel:epl11}  It is remarkable that information about the excited states of a non-trivial quantum many-body system can be obtained from the reduced density matrix, which is constructed solely from the ground-state wavefunction.\cite{Li:prl08,Sterdyniak:11}  As the ground-state can be more easily obtained than excited states via available numerical methods, understanding the implications of the ground-state quantum entanglement for the excited states is a central issue in quantum many-body theory, particularly in those systems with some type of topological order.\cite{Qi:rmp11,Nayak:rmp08,Wen}

The entanglement spectrum has been studied extensively in the fractional quantum Hall effect,\cite{Li:prl08,Regnault:prl09,Zozulya:prb09,Lauchli:prl10,Thomale_AC:prl10,Sterdyniak:prl11,Thomale:prb11} spin chains,\cite{Pollmann:prb10,Thomale:prl10,Poilblanc:prl10,Lauchli:prb12}, other one-dimensional systems,\cite{Turner:prb11,Deng:prb11,KK_negative_sign} topological superconductors,\cite{Fidkowski:prl10,Dubail:prl11} and topological insulators.\cite{Turner:prb10,Hughes:prb10,Kargarian:prb10,Prodan:prl10,Alexandradinata:prb11,Fiete:pe11} The information contained in the entanglement spectrum depends crucially on how one partitions (cuts) the Hilbert space of the system into two portions.  For example, a bond partition defined by a local cut across a bond of the one-dimensional Heisenberg spin-1/2 chain reveals an uninteresting structure, while partitioning the system in momentum space reveals the underlying U(1) conformal field theory (CFT) and allows a connection to be drawn to fractional quantum Hall systems.\cite{Thomale:prl10}  As a rule of thumb, one might expect local (in real-space) cuts to reveal more information in gapped systems, while non-local cuts (in real-space) may contain important information in gapless systems.\footnote{We thank R. Thomale for discussions on this point.}  

The one-dimensional Kugel-Khomskii model with $L$ sites,\cite{TMO_book}
\begin{equation}
\label{eq:KK}
H(X,Y)= \! \sum_{\ i=1}^L (\vec{S_i}\cdot\vec{S}_{i+1}+X)(\vec{\tau_i}\cdot\vec{\tau}_{i+1}+Y),
\end{equation}
contains both gapped and gapless regions in its phase diagram (as a function of $X$ and $Y$),\cite{Yamashita:prb98,Li:prl98,Pati:prl98,Azaria:prl99,Yasufumi:jpsj00,Itoi:prb00,Azaria:prb00,Zheng:prb01} so we will make use of both local and non-local cuts (in real-space) of the Hilbert space to obtain a comprehensive picture of its entanglement properties. In Eq.\eqref{eq:KK}, $\vec S_i$ represents the spin-1/2 degree of freedom on site $i$, and $\vec \tau_i$ is a two-state orbital degree of freedom on the same site. For general $X$ and $Y$, Eq.\eqref{eq:KK} possesses SU(2)$\times$SU(2) symmetry, with a larger SU(4) symmetry realized for $X=Y=1/4$.  We note that Eq.\eqref{eq:KK} admits an exact solution by Bethe Ansatz at the SU(4) symmetric point, where it is gapless,\cite{Sutherland:prb75} and at the point $X=Y=3/4$, where it is gapped\cite{Kolezhuk:prl98} and a topological ``non-Haldane" phase is realized.\cite{Nersesyan:prl97}  Phase diagrams for this model have been obtained via numerical methods and with field theory approaches.\cite{Yamashita:prb98,Li:prl98,Pati:prl98,Azaria:prl99,Yasufumi:jpsj00,Itoi:prb00,Azaria:prb00,Zheng:prb01}

In this paper, we investigate the entanglement properties of the one-dimensional Kugel-Khomskii model \eqref{eq:KK} via exact diagonalization and analytical techniques.  We begin with a study of the spin-orbital entanglement entropy using larger system sizes ($L=16$) than previously reported ($L=8$).\cite{Chen:prb07}  For 8 sites, we find that we are able to reproduce the results of Ref.[\onlinecite{Chen:prb07}], but we disagree with their physical interpretation.   With 12 and 16 site lattices we obtain a better understanding of finite size effects present for $L=8$ and find a phase diagram in overall agreement  with other studies.\cite{Yamashita:prb98,Li:prl98,Pati:prl98,Azaria:prl99,Yasufumi:jpsj00,Itoi:prb00,Azaria:prb00,Zheng:prb01}  

In the remainder of the paper, we will describe the results of our study of the entanglement spectrum of the one-dimensional Kugel-Khomskii ground state with a focus on its antiferromagnetic phases. A central goal of this work is to understand how a ground state wavefunction manifests its physical properties through the entanglement spectrum derived from a reduced density matrix. A density matrix is formed by tracing over degrees of freedom from the pure ground state density matrix. The degrees of freedom to be traced over are defined by a partition of the system in two parts. These partitions are loosely defined by cuts which represent artificial boundaries between the two partitions. The different cuts that we have considered are the orbital or ``rung" cut, the bond or ``leg" cut and the momentum cut. The orbital cut is defined by tracing over the orbital degrees of freedom from all the sites of the chain. The bond cut is defined by cutting the along two bonds of the periodic chain, leaving only half of the sites untraced. The momentum cut is non-local in real space and utilizes a mapping to a magnon representation of the ground state.    
  
One of our main results is that we find two ``entanglement gaps" with the rung cut. One gap seems to be characteristic of the gapless antiferromagnetic region of the phase diagram and the other {\em separate} entanglement gap is characteristic of the gapped dimerized non-Haldane phase.  In both cases the low-lying entanglement spectrum is {\em not} related to the Hamiltonian of the ``virtual edges"\cite{Poilblanc:prl10} if the system is viewed as a two-leg ladder (which would be a spin-1/2 Heisenberg model).  This is intimately related to the fact that the phases of \eqref{eq:KK} are driven by the ``four-spin" interaction $(\vec{S_i}\cdot\vec{S}_{i+1})( \vec{\tau_i}\cdot\vec{\tau}_{i+1})$ that pushes physics away from either the rung singlet or Haldane phase favored by a ``rung coupling" $\vec S_i \cdot \vec \tau_i$.\cite{Nersesyan:prl97,Hijii:prb09}  Both the rung singlet and Haldane phase exhibit low-lying entanglement eigenvalues that mimic the physical excitations of a virtual edge.\cite{Poilblanc:prl10,Lauchli:prb12}  In addition, we have generalized the non-local momentum cut of Thomale et. al.\cite{Thomale:prl10} to spin chains with orbital degeneracy. However, we \emph{do not} find an entanglement gap in the entanglement spectrum at the gapless SU(4) point with this cut. We observe a new counting of the low-lying levels that has yet to be matched with expectations of the corresponding SU(4)$_1$ Wess-Zumino-Witten conformal field theory as would be anticipated from the study of U(1) conformal field theories.\cite{Thomale:prl10}

Our paper is organized as follows. In Section \ref{sec:KK}, we present the phase diagram of the Kugel-Khomskii model Eq.\eqref{eq:KK}, and identify special points of interest that we will focus on in our entanglement study. In Section \ref{sec:RC}, we study the entanglement entropy and the entanglement spectrum of a partition between spin and orbital degrees of freedom. In Section \ref{sec:BC}, we look at the bond or leg cut entanglement spectrum. In Section \ref{sec:MC}, we consider a cut in momentum space, and in Section \ref{sec:conclusions} we present the main conclusions of our paper.  Several technical results related to the ``orbital" cut and the momentum cut are presented in the two Appendices.

\section{Phase Diagram of the Kugel-Khomskii Chain}
\label{sec:KK}
The one-dimensional Kugel-Khomskii model, Eq.\eqref{eq:KK}, is rich in both gapless and gapped phases with interesting symmetries.\cite{Yamashita:prb98,Li:prl98,Pati:prl98,Azaria:prl99,Yasufumi:jpsj00,Itoi:prb00,Azaria:prb00,Zheng:prb01} In one-dimension this model can be thought of as either a two-leg spin ladder (spin on one leg, orbitals on the other) or a one-dimensional spin-orbital chain at quarter filling. Physically, the Kugel-Khomskii model (in one, two, and three dimensions) naturally arises in the strong coupling limit (large Hubbard $U$ limit) of transition metal oxides,\cite{TMO_book} a class of materials which have recently been predicted to contain topological and other exotic magnetic phases under a variety of different conditions.\cite{Ruegg:prl12,Kargarian:prb11,Wan:prb11,Wan:prl12,Pesin:np10,Hu:prb13,Ruegg11_2,Ruegg:prb12,Yang:prb11,Yang:prb11,Yang:prb10,Shitade:prl09,Jackeli:prl09,Chaloupka:prl10,singh:prl12,Jiang:prb11,Xiao:nc11,Kargarian12}

\begin{figure}[h]
\includegraphics[width=7cm]{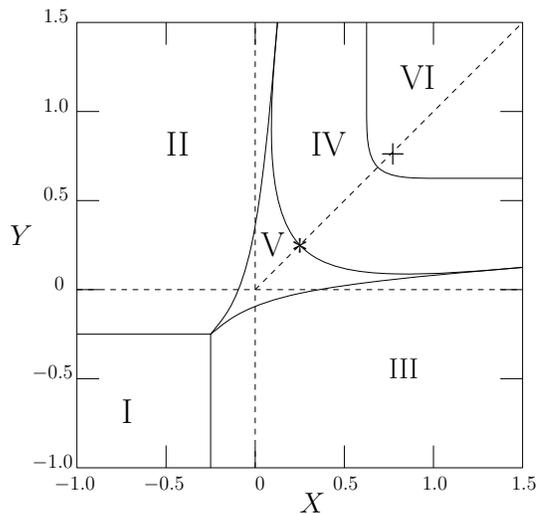}
\caption{The phase diagram of the one-dimensional Kugel-Khomskii model given in Eq.\eqref{eq:KK}. Phase I is a fully polarized ferromagnetic (FM) state for both spin and orbital sectors.  In phase II, the ground state is antiferromagnetic (AFM) for spin and FM for orbital degrees of freedom.  In phase III, spin and orbital orders are reversed relative to phase II by $X \leftrightarrow Y$. Phases IV and VI are gapped dimerized phases, and phase V is a gapless phase with non-trivial spin-orbital entanglement. The SU(4) point  at $X=Y=1/4$ is indicated by a {\bf *} and the exactly solvable point $X=Y=3/4$ indicated by a {\bf +} is in the non-Haldane phase with finite string order.}
\label{pd}
\end{figure}

The phase diagram of Eq.\eqref{eq:KK} has been studied with various field theoretical and numerical methods, \cite{Yamashita:prb98,Li:prl98,Pati:prl98,Azaria:prl99,Yasufumi:jpsj00,Itoi:prb00,Azaria:prb00,Zheng:prb01} and is presented in Fig. \ref{pd}.  Phase I is a fully polarized ferromagnetic (FM) state for both spin and orbital sectors.  In phase II, the ground state is antiferromagnetic (AFM) for spin and FM for orbital degrees of freedom.  In phase III, spin and orbital orders are reversed relative to phase II by $X \leftrightarrow Y$. Phases IV and VI are dimerized AFM phases, and phase V is a gapless phase AFM in spin and orbital sectors with non-trivial spin-orbital entanglement.  Thus, phases I, II, III, and V are gapless, while phases IV and VI are gapped. 

The lowest energy excitations of phase IV and VI differ in the character of their lowest excited states which are magnons. Despite that, phase IV and VI are really the same phase, which is the non-Haldane phase.\cite{Nersesyan:prl97} At the boundary line separating IV and VI in Fig.\ref{pd}, the ground state is in fact fully gapped\cite{Zheng:prb01,Yasufumi:jpsj00} and the transition between IV and VI should be interpreted as a smooth crossover.\cite{Azaria:prb00} This unfortunately has caused some confusion in the literature\cite{Chen:prb07} where a distinction has been made between these two phases. In fact, we will show that the entanglement properties of the ground state are smooth across the combined region of IV and VI, once subtleties due to degeneracies and finite size effects in the dimerized phase are taken into account. In this paper when we refer to phase IV, we will often mean both phase IV and VI, unless an explicit distinction is made. 

At $X=Y=1/4$ this model contains a global SU(4) symmetry, is gapless, and is exactly solvable by Bethe ansatz.\cite{Sutherland:prb75} This SU(4) symmetry point has central charge c=3 and is described by and SU(4)$_1$ Wess-Zumino-Witten(WZW) theory at low-energies, which establishes it as a phase that is not simply the ``sum" of two gapless legs of a spin-orbital ladder (as the latter would have central charge c=2), but rather the sum of two SU(2)$_2$ theories, each of which has central charge c=3/2.\cite{Itoi:prb00}  In the two-leg ladder picture, it is the ``four-spin" interaction $(\vec{S_i}\cdot\vec{S}_{i+1})( \vec{\tau_i}\cdot\vec{\tau}_{i+1})$ favoring dimerization that is responsible for much of the interesting physics of this model.\cite{Hijii:prb09,Nersesyan:prl97}

At $X=Y=3/4$ this model is gapped, but has exactly doubly degenerate ground states. As was shown by Kolezhuk and Mikeska,\cite{Kolezhuk:prl98} either ground state has an expression in the form of a matrix product state(MPS) and both spontaneously break the original translational symmetry of the chain by exhibiting an exact dimerization. In the two-leg ladder picture, these ground states are the two inequivalent exact staggered dimer coverings of the ladder. The phase at this Kolezhuk and Mikeska(KM) point where $X=Y=3/4$ is an example of a non-Haldane quantum spin liquid\cite{Nersesyan:prl97} with a non-local order parameter: it has a non-zero string order parameter\cite{denNijs:prb89,Kennedy:prb92} like the Haldane state,\cite{Haldane:prl83} but possess a finite dimer correlation that distinguishes it from the Haldane phase. 

\section{Orbital or Rung Cut}
\label{sec:RC}

Turning now to entanglement physics, we begin by exploring the properties of the reduced density matrix obtained by tracing out the orbital degrees of freedom.  If one interprets the Hamiltonian \eqref{eq:KK} in the language of spin chains, as a two-leg ladder with spin on one leg and orbital on the other, the reduced density matrix would be obtained by tracing out one leg. Such a partition of degrees of freedom corresponds to a cut through the rungs of the ladder. To make connection with previous work on two leg ladder spin chains, we will refer to this ``cut" of the Hilbert space as the ``rung cut". We will also sometimes refer to the orbital degrees of freedom as pseudospin degrees of freedom. Previously this rung cut was studied with exact diagonalization by Poilblanc,\cite{Poilblanc:prl10} and L\"auchli and Schliemann\cite{Lauchli:prb12} for a system with only two-spin interactions. Their results are therefore rather different from ours, where the ``four-spin" interaction is crucial to the physics.

\subsection{Entanglement Entropy}\label{sec:SEnt}
In this section, we present and analyze our numerical results based on exact diagonalization for the spin-orbital entanglement entropy $S_\text{ent}(X,Y)$ (obtained from the reduced density matrix, $\rho_{\rm red}$, with the ``rung cut") as a function of the model parameters $X$ and $Y$. The entanglement entropy, also known as the von-Neumann entropy, is obtained as
\begin{equation}
 S_\text{ent}(X,Y)=-\frac{1}{L}{\rm Tr}\bigl\{\rho_{\rm red}(X,Y)\ln[ \rho_{\rm red}(X,Y)]\bigr\},
 \end{equation}
where Tr denotes the trace over the (real) spin degrees of freedom of the reduced density matrix and we have chosen to normalize by the chain length $L$. Strictly speaking $S_\text{ent}$ is an intensive quantity and should really be termed the entropy density. However, for convenience and comparative purposes, we will just refer to $S_\text{ent}$ as the entropy. In all our numerical calculations we use periodic boundary conditions. To avoid ground state degeneracies in region V, we limit ourselves to chain lengths $L$ of multiples of $4$. System sizes of $L=4n+2$ with $n$ an integer possess additional degeneracies at the SU(4) point, that do not reflect the behavior in the thermodynamic limit, so we do not consider those system sizes.\cite{Yamashita:prb98} 

The entanglement entropy obtained from the orbital or rung cut has been studied before in the pioneering work of Chen {\it et al.} \cite{Chen:prb07} for $L$ up to 12 (though only data and figures are presented for $L=8$).  Based on the system sizes they studied, the authors of Ref.[\onlinecite{Chen:prb07}] concluded that $S_\text{ent}(X,Y)$ can reveal phase boundaries between the various ground states in Fig.\ref{pd} and provide a finer characterization within the ground state phases earlier found by Itoi {\it et al.}\cite{Itoi:prb00} A key result of Chen {\it et al.} is that $S_\text{ent}(X,Y)$ reproduces the phase diagram of Zheng and Oitmaa.\cite{Zheng:prb01}

In particular, Chen {\it et al.}\cite{Chen:prb07} argued (based on the entanglement entropy) that the transitions between regions I, II and III with V in Fig.\ref{pd} are first order.  The first order nature of these transitions is manifested as a large discontinuity in $S_\text{ent}$ from zero (in phases I, II, and III) to a intermediate value (for region V).\footnote{The scale of the von-Neumann entropy for a orbital cut partition of a two leg spin-ladder is set by the upper bound $\log_\mathrm{e}2 \approx 0.693$ which is saturated by a product state of spin-orbit singlets.} This result is upheld by our study, and others. The boundary between regions VI and IV in Fig.\ref{pd} also appears as a discontinuity in $S_\text{ent}$, but to a lesser degree.  However, in Ref.[\onlinecite{Chen:prb07}] it was not appreciated that both region VI and IV are in the same phase and are only distinguished by the nature of their lowest lying (gapped) excitations which are magnons.\cite{Azaria:prb00,Zheng:prb01} In fact, the ``phase boundary" separating them has only gapped excitations and thus their ground states are adiabatically connected. Hence, it is surprising to find such a large quantitative change in $S_\text{ent}$. The more severe transition between the gapped and gapless phase which corresponds to the boundary separating regions V and VI was identified with ``ridge lines" in the topography of $S_\text{ent}(X,Y)$,\cite{Chen:prb07} where a ridge is taken to mean a line of points where there exists at least one direction in which $S_\text{ent}$ is a local maximum.  It was then claimed that the ridges meet at the SU(4) symmetry point, $X=Y=1/4$, which was also claimed to be a distinguished local maximum.

In this early study of the von-Neumann entropy of the spin-orbital entanglement in the Kugel-Khomskii chain, exact diagonalization data was reported almost entirely for chains of lengths $L=8$ and to a limited degree $L=12$.\cite{Chen:prb07}  In our study we have reproduced their data for $L=8$ chains and also extended our study to include $L=12$ and $L=16$ chains. Our results on larger system sizes show some of their conclusions are incorrect and their misidentified phase boundaries are due to level crossings between almost degenerate ground states and finite size effects. Degenerate ground states in phase IV and VI are a symptom of the spontaneously broken symmetry of the non-Haldane phase caused by dimerization. However this degeneracy is only approximate at small system sizes and level crossings between these ground states can have significant effects on the von-Neumann entropy. We shall argue that with the exception of the phase boundaries between regions I,II, and III with V, $S_\text{ent}$ by itself cannot sharply reveal many of the phase boundaries with finite system size studies. Instead, the entanglement spectrum proves to be a more powerful tool by revealing more structure.

For completeness, we briefly review our numerical procedures. To compute the ground state via exact diagonalization we consider the $\sum_i S^z_i=\sum_i \tau^z_i=0$ sector. We make use of the fact that linear momentum is a conserved quantum number and will take the value of either $\pi$ or $0$ due to time reversal symmetry. In the study of $L=8$ chains, we have performed exact diagonalizations in a grid of points spaced $\Delta X=\Delta Y = 0.005$ apart. In our $L=12$ study, because of computing constraints we were limited only to a resolution of $\Delta X=\Delta Y = 0.01$.  In our work we focus only in the region $0\leq X,Y \leq 1$, or regions II, III, IV, V and VI of Fig.\ref{pd}.

\begin{figure*}
\centering
\includegraphics[width=2\columnwidth]{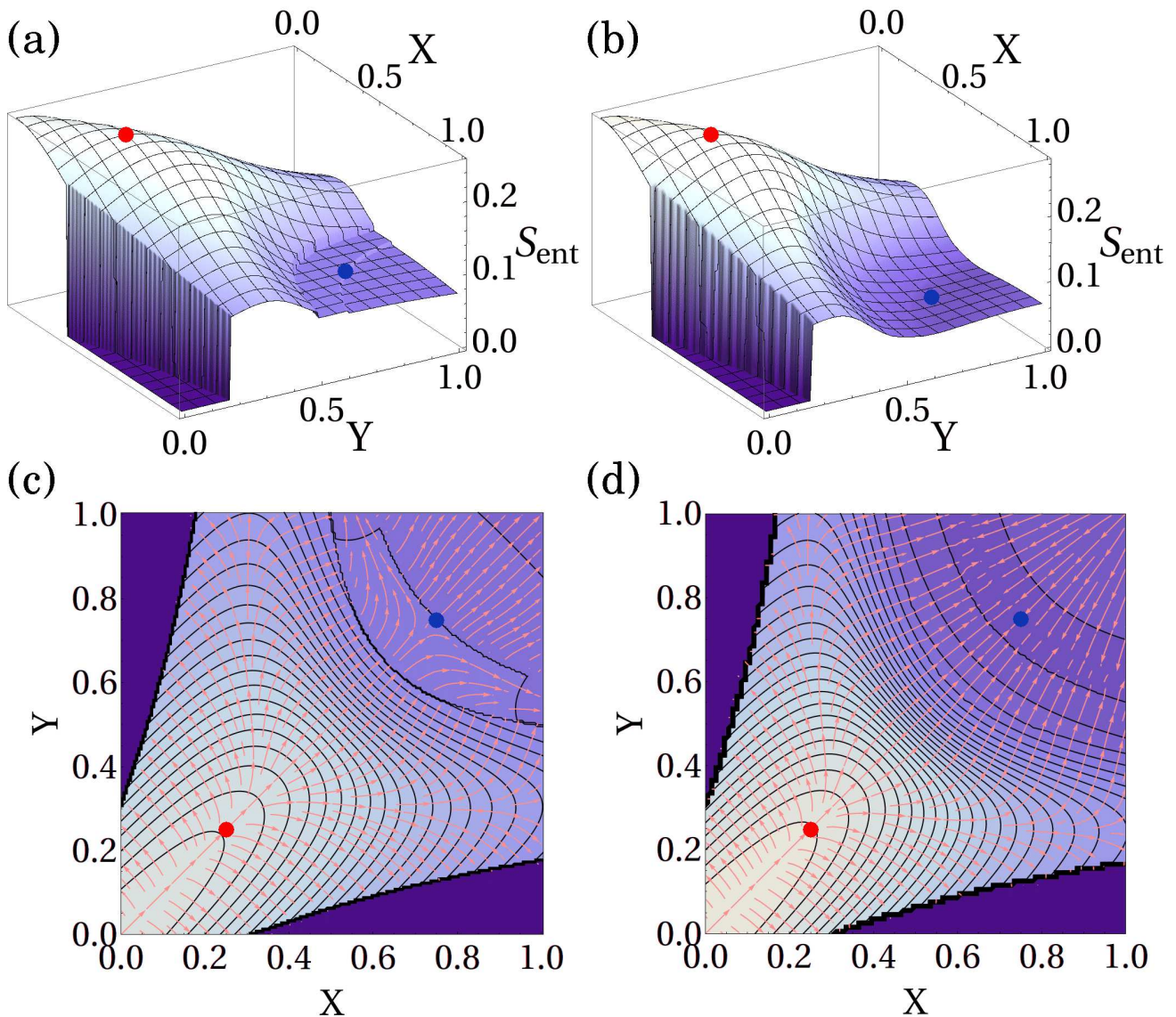}
\caption{(color online) The spin-orbit entanglement entropy $S_\text{ent}(X,Y)$. Surface plots of $S_\text{ent}(X,Y)$ for (a) $L=8$ and (b) $L=12$ system sizes. The red and blue dots are indicative of the SU(4) symmetry ($X=Y=1/4$) point and the Kolezhuk-Mikeska(KM) point ($X=Y=3/4$), respectively.  Note the two lines of discontinuous jumps in (a), in which the KM point (blue dot) lies on one of these lines. Contour plots for (c) $L=8$ and (d) $L=12$. The pink arrows represent streamlines in the direction of steepest descent.}\label{SEnt}
\end{figure*}

In Fig.\ref{SEnt}(a) we reproduce the main result of Chen {\it et al.}\cite{Chen:prb07} but with much greater resolution. The line of discontinuities in $S_\text{ent}$, which was thought to separate regions IV and VI, is now made more apparent. However, there is also a second line of discontinuities which is revealed and which contains the exact staggered dimer KM point. This second line was not interpreted to correspond to a phase boundary previously. However, we identify both these lines of discontinuities as corresponding to level crossings between the total momentum $K=0$ and $K=\pi$ ground states in the gapped phase. Field theory studies predict a spontaneous dimerization in the gapped phase which requires at the very least, doubly degenerate ground states.\cite{Nersesyan:prl97} But with finite system sizes, this degeneracy is lifted and the eigenstates are only distinguished by their linear momentum, in this case $K=0$ and $K=\pi$. Previous numerical studies\cite{Yasufumi:jpsj00} have revealed these almost degenerate lowest eigenstates in this region of the phase diagram. But with increasing system size we expect these artifacts to diminish in severity and this is indeed what happens with $L=12$.

\begin{figure}[th]
\centering
\includegraphics[width=0.7\columnwidth]{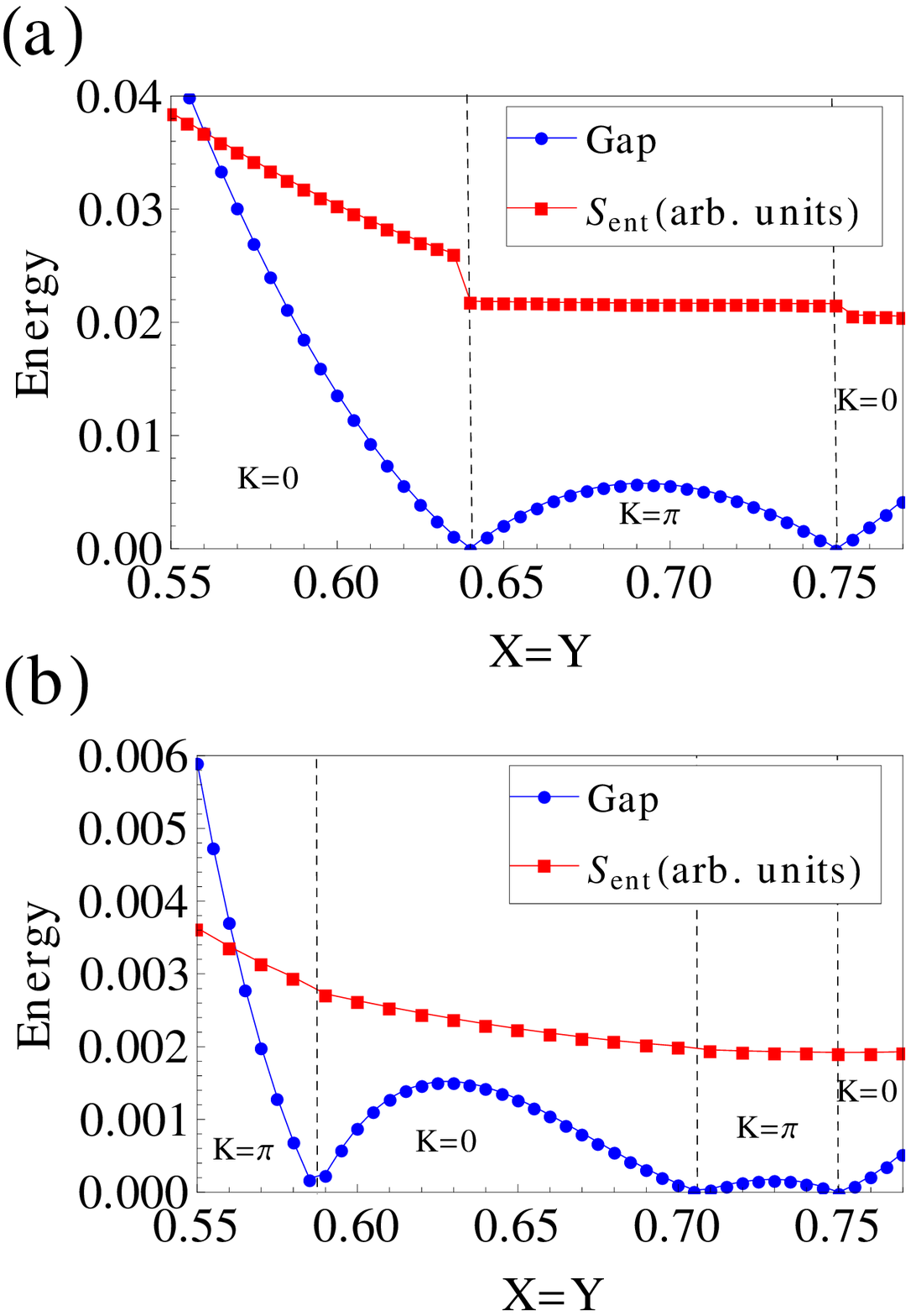}
\caption{(color online) (a) The excitation gap between the two lowest eigenstates for the $L=8$ chain in the $\sum_i S^z_i=\sum_i \tau^z_i=0$ sector along the $X=Y$ line in the region of level crossings. Shown for qualitative comparison is $S_\text{ent}$ in scaled units. In regions where the gap is non-zero, the linear momenta of the absolute ground state is also labeled. (b) The analogous result for $L=12$ chains.}\label{gaps}
\end{figure}

Shown in Fig.\ref{SEnt}(b) is a surface plot of $S_\text{ent}(X,Y)$ corresponding to chains of length $L=12$, where $S_\text{ent}(X,Y)$ is now continuous as a function of $X$ and $Y$ with the exception of the boundaries with the ferromagnetic phases, II and III (in the orbital sector and spin sector, respectively). In place of the above mentioned two lines of discontinuity for $L=8$, are kinks with a discontinuous first derivative. Moreover, the position of these lines of kinks are shifted from that of the discontinuous jumps seen in the $L=8$ study and there are more kinks in the $L=12$ case. This can also be clearly seen along the $X=Y$ line shown in Fig.\ref{gaps}. Thus, we see that there are significant qualitative and quantitative differences with the smaller ($L=8$) system size.

Shown in Figs.\ref{SEnt}(c-d) are contour plots of $S_\text{ent}(X,Y)$ with arrows drawn which represent the streamlines in the direction of steepest descent. These lines allow one to estimate the position of the ridge lines which correspond to the separatrices of the streamlines. One ridge line extends along the $X=Y$ diagonal and two others lie almost parallel to the $X$ and $Y$ axes. Also highlighted are the SU(4) and exact staggered dimer KM point, indicated with a red and blue dot, respectively. In particular, the SU(4) symmetry point does not seem to be a distinguished local maximum from the other points on the $X=Y$ line in either the $L=8$ or the $L=12$ system. If one views the contours in the increasing $X=Y$ direction, local peaks and valleys are recognized as concave and convex contours respectively. The point where the contours change their curvature is according to Ref.[\onlinecite{Chen:prb07}] where the SU(4) point should lie. However, the SU(4) point as marked in Fig.\ref{SEnt}(c,d) does not lie on such a point. Furthermore, the point at which the ridge lines meet lies beyond $X=Y=1/4$ in the increasing $X$ and $Y$ direction. These observations are in contradiction with the conclusions of Chen {\it et al.}\cite{Chen:prb07}

The level crossings in regions IV and VI may be identified by a closure of the energy excitation gap between the two lowest eigenstates in the $\sum_i S^z_i=\sum_i \tau^z_i=0$ sector. In Fig.\ref{gaps} we present plots of these energy gaps along the $X=Y$ line in the region of level crossings. Also shown are $S_\text{ent}(X,Y)$ curves. It is clear the positions of the level crossings match the discontinuous jumps or kinks in $S_\text{ent}(X,Y)$. However, the position of the kinks for the $L=12$ data are harder to distinguish. Nevertheless, we can locate these points from the positions where the second derivative $S_\text{ent}$ is peaked and they do agree with the positions of the gap closures. Thus we can confirm that indeed the large qualitative changes in the entanglement entropy can be attributed to levels crossing between almost degenerate ground states. Moreover, each level crossing is always accompanied by a change in the time-reversal invariant linear momentum.

\begin{figure}[th]
\centering
\includegraphics[width=0.7\columnwidth]{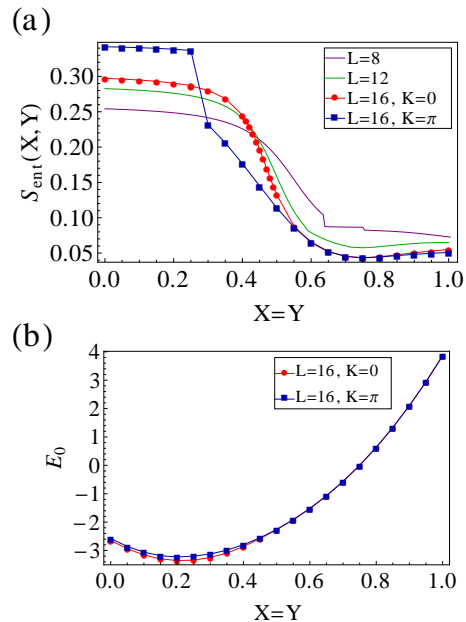}
\caption{(color online) (a) $S_\text{ent}(X,Y)$ along the $X=Y$ line for chain lengths $L=8,12,16$. The data presented for $L=16$ includes ground states from both time-reversal invariant linear momenta $K=0,\pi$. (b) The ground state energy $E_0$ of $L=16$ chains from the relevant linear momenta sectors.}
\label{Sent_E0}
\end{figure}

In Fig.\ref{Sent_E0}(a) traces of $S_\text{ent}(X,Y)$ along the $X=Y$ line for all chain lengths we studied are presented and compared. With increasing chain length, the regions corresponding the to gapless phase V have ever increasing spin-orbital entanglement entropy. Note that in this region the $L=16$ ground states are non-degenerate and have $K=0$ as demonstrated in Fig.\ref{Sent_E0}(b) where the relevant ground state energies are plotted. Conversely, the gapped regions IV and VI exhibit ever decreasing entanglement entropy with increasing system size. In this region, the differences between $L=16, K=0$ and $L=16, K=\pi$ are small, both in their ground state energy, $E_0$, and their entanglement entropy, $S_\text{ent}$. This is again consistent with the expectation that with increasing system size the numerical eigenstates tend to exact degeneracy in the gapped phase.

Since in the thermodynamic limit of the gapped, dimerized non-Haldane phase, the two lowest eigenstates must be doubly degenerate, we do not expect the kinks or jumps in $S_\text{ent}$ to persist in the infinite size limit unless it involves a transitions between the AFM and FM phase. This implies a weakness in determining phase boundaries by solely relying on large qualitative changes in the entanglement entropy and the subtleties involved with almost degenerate ground states. In the next section we examine the entanglement spectrum, which allows one to gain more insight into the nature of the ground state phases and the transitions between them.

\subsection{Orbital or Rung Cut Entanglement Spectrum}

The entanglement spectrum(ES) is the spectrum of an  ``entanglement Hamiltonian", $H_{\rm ent}$, obtained from the reduced density matrix via the relation,\cite{Li:prl08}
\begin{equation}
\rho_{\rm red}={\rm exp}\left\{-H_{\rm ent}\right\},\label{es}
\end{equation}
and we denote the $n^{th}$ eigenvalue of $H_{\rm ent}$ by $\xi_n$. One may think of the ES as just the eigenvalues of the reduced density matrix arranged on a negative logarithmic scale. There is in fact a simple relationship between the ES and $S_\text{ent}$ given by
\begin{equation}
S_\text{ent}=\frac{1}{L}\sum_n \xi_n \exp(-\xi_n),
\end{equation}
where the sum runs over the entire ES. 

In the remainder of this section, we will study the ES derived from the orbital or rung cut and plot the spectrum as a function of the momentum of the sub-system, which is a good quantum number due to the translational symmetry of the cut. In later sections, we will also study the entanglement spectrum for a bond cut, and a cut in momentum space.

In our study of the rung cut ES, we will consider first the effects of symmetry on the ES and discuss the structure of generic spectra in the $(X,Y)$ phase diagram. Then we will specialize to the case where there is additional $\mathbb{Z}_2$ symmetry on the $X=Y$ line.  Next we will focus on a few special points on this line, such as the exactly solvable KM point at $X=Y=3/4$ and the SU(4) symmetry point at $X=Y=1/4$. We will study a few key slices of parameter space, such as along the line $0\leq X=Y \leq 1$ and $X+Y=0.4$ for $0\leq X\leq 0.4$ as well. The latter line starts in region II of Fig.\ref{pd}, and continues through region V before ending in region III. Then, we will study the correlation functions derived from the Schmidt vectors ({\it i.e.}, the eigenvectors of $H_{\rm ent}$) in the gapless phase V, in particular at the SU(4) point. Lastly, we will study the changes to the ES when an exchange or rung coupling between local orbital and spin is introduced.

\subsubsection{Generic Symmetries of the Entanglement Spectrum}\label{sec:symm}

We first establish the notation for our discussion on the ES. Let $\mathcal{H}_{1/2}$ denote the Hilbert space of a chain of $L$ spin-1/2 moments. To make connection with previous studies\cite{Nersesyan:prl97,Li:prl98,Kolezhuk:prl98,Poilblanc:prl10} we will for convenience take the two leg ladder interpretation of the spin-orbital Kugel-Khomskii chain, Eq.(\ref{eq:KK}).  From this perspective, the spin-1/2 chain with orbital degeneracy may be thought of as a two leg ladder system with one chain representing a collection of $L$ spin degrees of freedom ($\mathcal{H}_S$) and the other chain representing $L$ orbital or pseudospin degrees of freedom ($\mathcal{H}_\tau$). The total combined Hilbert space is then $\mathcal{H}_{S\tau}:=\mathcal{H}_S\otimes \mathcal{H}_\tau$. Thus $\rho_\text{red}[\Psi] = \text{Tr}_{\mathcal{H}_\tau}| \Psi\rangle \langle \Psi|$ is the reduced density matrix of an orbital or rung cut for the ground state (GS) $|\Psi \rangle$.

Symmetries of the GS $|\Psi\rangle$ implies certain symmetries of the ES. It should be noted, however, that a symmetry of $|\Psi \rangle$ is not necessarily a symmetry of the Hamiltonian $H(X,Y)$, but in the cases where $|\Psi\rangle$ is non-degenerate, a symmetry of $H(X,Y)$ turns out to be a symmetry of $|\Psi\rangle$ modulo a phase. We shall discuss the effects of each symmetry on the ES. 

Firstly, translational symmetry allows $\rho_\text{red}[\Psi]$ to be block-diagonalized by crystal momentum quantum numbers if $|\Psi\rangle$ itself is a momentum eigenstate. Hence the ES may be resolved according to momentum to reveal interesting structure. In fact, the ES of quantum spin-ladders\cite{Poilblanc:prl10} and Projected Entangled Pair states (PEPs) on cylindrical geometries\cite{Cirac:prb11} had previously been studied in this manner. 

The block diagonal structure is most easily revealed by expanding $|\Psi\rangle$ in a Fourier basis of $\mathcal{H}_{1/2}(\cong \mathcal{H}_S,\mathcal{H}_\tau)$. The allowed momenta are $p=2\pi\frac{n}{L}$ with $n=-L/2+1,\dots,L/2$. We denote such a basis by $\{p\alpha_p\}$ where $\{\alpha_p\}$ for fixed $p$ is an orthonormal basis for the momentum subspace $p$. With this basis we can expand any wavefunction of $\mathcal{H}_{S\tau}$ with the tensored basis $\{|p\alpha_p\rangle \otimes |q\beta_q\rangle\}$. 

Now let $|\Psi\rangle$ be a ground state of (\ref{eq:KK}) with momentum $K$. Expanding in the  Fourier basis gives 
\begin{align}
\left| \Psi \right\rangle = \sum_{\substack{p+q=K \\ \text{mod}\,2\pi}} \sum_{\alpha_p,\beta_q} \Psi_{p\alpha_p;q\beta_q} \left| p\alpha_p \right\rangle \otimes \left| q \beta_q \right\rangle. 
\end{align}
Forming $|\Psi\rangle \langle \Psi | $ and then taking the trace over $\mathcal{H}_\tau$ yields a block diagonalized reduced density matrix,
\begin{align}
&\rho_{\text{red}}[\Psi] \nonumber \\  
&= \sum_{\substack{p\\ \alpha_p,\alpha_p'}}\left\lbrace\sum_{\beta_q} \left( \Psi_{p\alpha_p;q\beta_q}\right) \left( \Psi^{*}_{p\alpha_p';q\beta_q}\right) \right\rbrace \left|  p\alpha_p \right\rangle \left\langle p \alpha_p' \right|, 
\end{align}
with $q=K-p$.

Strictly speaking, the dimerized ground states spontaneously break translational symmetry and thus cannot have their ES momentum resolved. But symmetric and anti-symmetric superpositions of such degenerate symmetry broken states are momentum eigenstates with momentum $K=0$ and $K=\pi$ respectively. In a later section that analyzes the entanglement spectrum of the Kolezhuk-Mikeska(KM) point ($X=Y=3/4$) these subtleties will be discussed in detail. Nevertheless, in our finite size exact diagonalization studies, the degeneracy in the AFM gapped phase IV is often inexact except perhaps at level crossings between $K=0$ and $K=\pi$ ground states as was discussed in Sec.\ref{sec:SEnt}. This was observed in Fig.\ref{SEnt} as lines of discontinuous jumps or kinks in $S_\text{ent}$ which also contained the KM point(blue dot). Nevertheless, generically the non-degenerate numerical ground states obtained by exact diagonalization in the gapped AFM region are one of the two time-reversal momentum groundstates with a small but finite gap between their energies. 

The ES can also be read off from a Schmidt decomposition (SD) of $|\Psi\rangle$ with respect to the partition between $\mathcal{H}_S$ and $\mathcal{H}_\tau$. When the translational symmetry is also taken into account, the SD can take the following form

\begin{align}
|\Psi\rangle=\sum_p\sum_{\phi_p}\mathrm{e}^{-\xi_{\phi_p}/2}|p\phi_p\rangle \otimes |(K-p)\overline{\phi_p}\rangle, 
\label{eqn:SD}\end{align} 
where $K=0,\pi$ due to time reversal symmetry. $\{\phi_p\}$ and their duals $\{\overline{\phi_p}\}$ are the Schmidt vectors in the momentum sector $p$ and $(K-p)$ of $\mathcal{H}_S$ and $\mathcal{H}_\tau$ respectively. Moreover, they each independently form an orthonormal basis. Here $\xi_{\phi_p}$ is the entanglement eigenvalue that corresponds to $\phi_p$ and we denote the full ES by the set of pairs $\{(p,\xi_{\phi_p})\}$ which runs over all momenta and Schmidt vectors. For a fixed $|\Psi\rangle$ and $p$, a given $\phi_p$ uniquely determines $\overline{\phi_p}$. However, the above decomposition is not unique because of the phase degree of freedom in defining $\phi_p$ and $\overline{\phi_p}$.  For example the transformation $\phi_p\rightarrow\mathrm{e}^{i\theta}\phi_p$ and $\overline{\phi_p}\rightarrow\mathrm{e}^{-i\theta}\overline{\phi_p}$ leaves the SD invariant. Thus, the bar over $\overline{\phi_p}$ should not be mistaken for complex conjugation. 

The next crucial symmetry of $|\Psi\rangle$ is spatial inversion ($\mathcal{I}$) that maps sites $i=1,\dots,L$ according to  $i\mapsto -i\;\text{mod}\; L$. This is a unitary transformation and in momentum space this maps $p\mapsto -p$ mod $2\pi$. This symmetry will lead to a reflection symmetry in the momentum resolved ES such that $\{(p,\xi_{\phi_p})\}\equiv \{(-p,\xi_{\phi_{p}})\}$. To see this, note that for a non-degenerate GS $|\Psi\rangle$, it must be that $\mathcal{I}|\Psi\rangle=(-1)^{n_\mathcal{I}}|\Psi\rangle$ where $n_\mathcal{I}=0,1$. This follows from $\mathcal{I}^2=1$ and the non-degeneracy assumption. Next applying $\mathcal{I}$ to the SD of $\Psi$ (\ref{eqn:SD}) yields

\begin{eqnarray}
\mathcal{I}|\Psi\rangle &&=\sum_p\sum_{\phi_p}\mathrm{e}^{-\xi_{\phi_p}/2}|(-p)\phi_p\rangle \otimes |(K+p)\overline{\phi_p}\rangle \nonumber \\
&&=  (-1)^{n_\mathcal{I}} |\Psi\rangle. \nonumber \\ 
\end{eqnarray}           
Which implies 
\begin{eqnarray}
|\Psi\rangle &&= \sum_p\sum_{\phi_{p}}\mathrm{e}^{-\xi_{\phi_{p}}/2} (-1)^{n_\mathcal{I}} |(-p)\phi_{p}\rangle \otimes |(K+p)\overline{\phi_{p}}\rangle. \nonumber \\
\end{eqnarray}           
Next we compare $\rho_\text{red}[\Psi]$ and $\rho_\text{red}[(-1)^{n_\mathcal{I}}\mathcal{I}\Psi]$ which are two alternate expressions for the reduced density matrix
\begin{eqnarray}
\rho_\text{red}[\Psi] &&= \sum_p\sum_{\phi_p} \mathrm{e}^{-\xi_{\phi_p}}|p\phi_p\rangle \langle p\phi_p| \nonumber \\
\nonumber \\
\rho_\text{red}[(-1)^{n_\mathcal{I}}\mathcal{I}\Psi] &&= \sum_p \sum_{\phi_p} \mathrm{e}^{-\xi_{\phi_p}}|(-p)\phi_p\rangle \langle (-p) \phi_p|.\quad
\end{eqnarray}
Hence the spectrum of entanglement energy levels at $p$, $\{\xi_{\phi_p}\}$ must be identical to the spectrum at $-p$, $\{\xi_{\phi_{-p}}\}$. Thus we have a $p\rightarrow-p$ symmetry in the entanglement spectrum for non-degenerate groundstates. We will denote equivalences of spectra by the following suggestive notation, 
\begin{align}
\{(p,\xi_{\phi_p})\}\equiv \{(-p,\xi_{\phi_{p}})\} \equiv \{(p,\xi_{\phi_{-p}})\}
\end{align}
where $p$ ranges over all momenta and $\phi_p$ over all the associated Schmidt vectors. Note that non-degeneracy played a crucial role in ensuring that $\mathcal{I}$ maps a ground state back to itself modulo an overall sign. 

Next, due to the SU(2)$\times$ SU(2) spin and pseudospin (orbital) rotational symmetry, a GS $|\Psi\rangle$ if non-degenerate must necessarily be a singlet in both spin and pseudospin. That is $S^2_\text{tot}=\tau^2_\text{tot}=0$. Tracing over $\mathcal{H}_\tau$ keeps $\rho_\text{red}$ in the singlet $S^2_\text{tot}=0$ sector of $\mathcal{H}_S$. Thus, the momentum resolved entanglement eigenvalues of $\rho_\text{red}$ are expected to be generically non-degenerate unless additional symmetries or accidental degeneracies dictate otherwise. That is, there are no degeneracies to be associated with spin and pseudospin rotations. This should be contrasted with the ES derived from the gapped two-leg Heisenberg ladder.\cite{Poilblanc:prl10} There the GS instead satisfies the less restrictive $(\vec{S}_\text{tot}+\vec{\tau}_\text{tot})^2=0$ condition. This then leads to the ES having support in more than one $S^2_\text{tot}$ sector of $\mathcal{H}_S$. Here, the orbital $\vec{\tau}$-chain is identified with the second leg of the ladder which is traced over.

Lastly, the $S^2_\text{tot}=0$ condition dictates the maximum total number of physically significant entanglement energy levels from our numerically computed spectra. Let $N_\text{singlet}$ denote the dimension of the $S^2_\text{tot}=0$ subspace of $\mathcal{H}_S$ which grows with $L$. $N_\text{singlet}$ may be calculated by counting the number of standard Young tableaux with shape $[L/2,L/2]$ or 2 rows and $L/2$ columns. The ES reveals its physically relevant structure with only the lowest $N_\text{singlet}$ levels. In practice our computations generate more eigenvalues than $N_\text{singlet}$ because we work in the more convenient $S^z_\text{tot}=\tau^z_\text{tot}=0$ basis. These additional eigenvalues of $\rho_\text{red}$ should be zero in value based on physical grounds. But in our numerics they are seen to be finite but exponentially small compared the the rest of the spectrum.

\subsubsection{Generic Entanglement Spectra}\label{sec:ES_generic}

\begin{figure}
\centering
\includegraphics[width=\columnwidth]{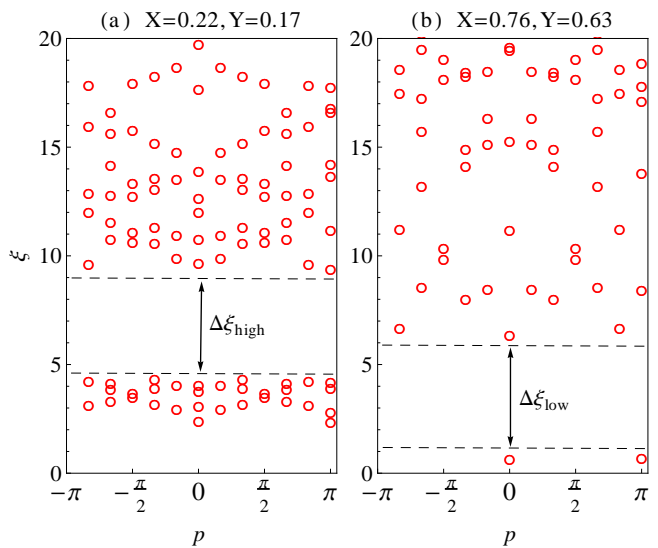}
\caption{(color online) Orbital or rung cut entanglement spectra against linear momentum $p$ at generic points in the AFM phases for $L=12$. The {\color{red}$\circ$} symbols denote singlet levels. (a) $(X,Y)=(0.22,0.17)$ from the gapless V phase. There are 32 low energy states across all $p$ that are separated by the gap $\Delta\xi_\text{high}$ from a continuum of high energy states. (b) $(X,Y)=(0.76,0.63)$ from the gapped VI phase. There are only 2 low energy levels at $p=0,\pi$ which are separated by a gap $\Delta\xi_\text{low}$ from a continuum of high energy states. The position of $\Delta\xi_\text{high}$ is above that of $\Delta\xi_\text{low}$.}\label{Fig:ES_generic}
\end{figure}

Our calculations of the orbital or rung cut ES from exact diagonalization of generic points in the AFM region do indeed exhibit the symmetry properties of the last subsection. Shown in Fig.\ref{Fig:ES_generic} are entanglement spectra taken from a $L=12$ chain derived from ground states deep in the gapless and gapped AFM phases. Both spectra are $p \rightarrow -p$ symmetric and are entirely composed of non-degenerate singlets. However the ES is interesting in other ways. The striking feature is an entanglement gap across all momenta exhibited by both spectra. In the gapless case Fig.\ref{Fig:ES_generic}(a), this gap labeled by $\Delta\xi_\text{high}$ separates a set of 32 low lying energy levels from a continuum of high energy levels. While in Fig.\ref{Fig:ES_generic}(b) a different entanglement gap $\Delta\xi_\text{low}$ separates a set of high levels from only two distinguished lowest levels at $p=0,\pi$. The corresponding ES for $L=8$ ($L=16$) is qualitatively similar but with less (more) entanglement levels. In phase V, data from $L=8,12$ and $16$ reveal an empirical relation $2^{L/2-1}$ for the number of lowest lying levels which are distinguished by the entanglement gap $\Delta\xi_\text{high}$. In phase IV, there are still only two lowest levels at $p=0,\pi$. 

Remarkably the rung cut ES of Fig.\ref{Fig:ES_generic} does not in any way resemble the gapped two-leg Heisenberg spin-ladder,\cite{Poilblanc:prl10} even in the gapped IV phase. The ES does not seem to exhibit any similarities, quantitative or qualitative with the real spectrum of a single Heisenberg chain. More specifically, triplets and higher degenerate multiples are absent from the spectrum. This could be interpreted in two different ways. On the one, hand it could mean the failure of the conjecture that relates the entanglement spectrum to the real energy spectrum of the (virtual) edges.\cite{Poilblanc:prl10} On the the other hand, it could be that the conjecture still holds true but the boundary or edge Hamiltonian is not the single AFM Heisenberg chain as one would expect by decoupling the spin and orbital chains in (\ref{eq:KK}).
 
The lack of triplets and higher degenerate multiplets can be traced back to the restriction to the $S_\text{tot}^2=0$ sector of the entanglement eigenvectors of $\rho_\text{red}$ as was shown in the previous section with symmetry based arguments. Hence we believe the second scenario to be more likely and the special nature of the Hamiltonian with the four spin interaction is the cause for the complicated boundary Hamiltonian. 

\subsubsection{$\mathbb{Z}_2$ Symmetry of the Entanglement Spectrum}

In this subsection we consider the effects of the $\mathbb{Z}_2$ symmetry operator which exchanges spin ($S$) and orbital ($\tau$) degrees of freedom. We denote such a linear operator by $\mathcal{F}$ which acts on $\mathcal{H}_{S\tau}$ by
\begin{equation}
\mathcal{F}\left(|\alpha\rangle\otimes|\beta\rangle\right)=|\beta\rangle\otimes|\alpha\rangle,
\end{equation}
for all $\alpha\in\mathcal{H}_S$ and $\beta\in\mathcal{H}_\tau$. Moreover, $\mathcal{F}$ is unitary and $\mathcal{F}^2=1$. Effectively, $\mathcal{F}$ exchanges $X\leftrightarrow Y$ when acting upon $H(X,Y)$. That is $\mathcal{F}H(X,Y)\mathcal{F}=H(Y,X)$. Thus a ground state $|\Psi\rangle$ of $H(X,Y)$ is mapped to a ground state of $H(Y,X)$ under $\mathcal{F}$. Applying $\mathcal{F}$ to the Schmidt decomposition of $|\Psi\rangle$ in (\ref{eqn:SD}) yields the symmetry $S_\text{ent}(X,Y)=S_\text{ent}(Y,X)$ which was noted by Ref.[\onlinecite{Chen:prb07}]. 
\begin{figure}
\centering
\includegraphics[width=\columnwidth]{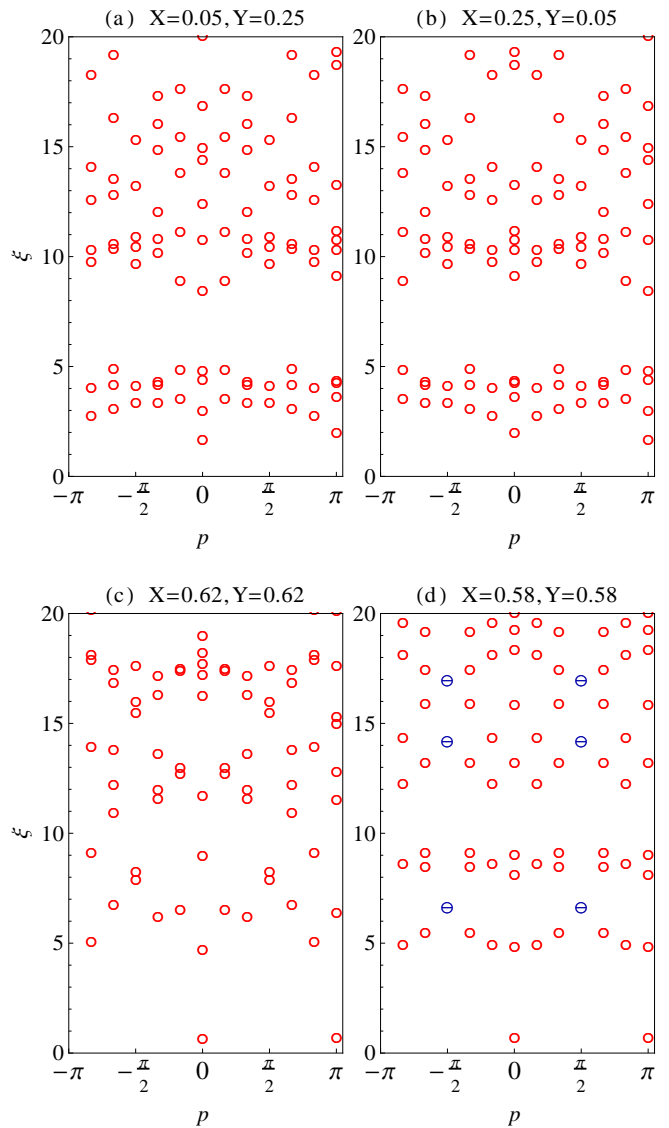}
\caption{(color online) Momentum resolved entanglement spectra for $L=12$ chains at points related by $\mathbb{Z}_2$ spin-orbital symmetry. The {\color{red}$\circ$} symbols denote singlet levels while the {\color{DarkBlue}\scriptsize$\ominus$} symbols denote doublets. (a) $(X,Y)=(0.05,0.25)$ in phase V. (b) $(X,Y)=(0.25,0.05)$ in phase V which is shifted by $p\rightarrow p+\pi$ relative to (a). (c) $(X,Y)=(0.62,0.62)$ in phase IV with ground state momentum $K=0$. (d) $(X,Y)=(0.58,0.58)$ in phase IV with $K=\pi$. Note the $\pi$ shift and doublets at $p=\pm\pi/2$ are present regardless of whether or not the ground state is in phase V or IV.}
\label{Fig:ES_Z2XY}
\end{figure}
However, it is {\em not} necessarily true that the ES derived from the GS of $H(X,Y)$ is identical to the one derived from $H(Y,X)$. It is only that true that both spectra produce the same von-Neumann entropy $S_\text{ent}$. In fact, using symmetry arguments similar to that of Section \ref{sec:symm}, it can be shown that when $K=\pi$ they are shifted by $\pi$ relative to one and another, otherwise they are identical when $K=0$. These arguments are presented in Appendix \ref{App:ES}. Shown in Fig.\ref{Fig:ES_Z2XY}(a-b) are entanglement spectra exhibiting this relative shift. 

Next, we specialize to the $\mathbb{Z}_2$ symmetric $X=Y$ line where $\mathcal{F}$ is a symmetry of $H(X,X)$. In this instance there may be double degenerate levels (doublets) present in the momentum resolved ES. These doublets occur at $p=\pm \pi/2$ but only when $K=\pi$. Shown in Fig.\ref{Fig:ES_Z2XY}(c-d) are entanglement spectra taken from phase IV on the $X=Y$ line. Fig.\ref{Fig:ES_Z2XY}(c) comes from a GS with $K=0$ whilst Fig.\ref{Fig:ES_Z2XY}(d) from one with $K=\pi$. Details of this argument for this degeneracy are also presented in Appendix \ref{App:ES} where it also requires that  $|\Psi\rangle$ satisfy $\mathcal{F}|\Psi\rangle = -|\Psi\rangle$. These results emphasize the importance the ground state momentum $K$ may have on the entanglement spectrum at finite systems sizes. It is also one more example of how the entanglement spectrum is {\em much more} sensitive to the details of the ground state than the von-Neumann entropy. 

In the next sections we will present our results and analyses on the entanglement spectra at the two special points along the $X=Y$ symmetric line. These are the SU(4) symmetric point ($X=Y=1/4$) and the exactly solvable Kolezhuk-Mikeska(KM) point ($X=Y=3/4$).  

\subsubsection{Entanglement Spectrum for the Exactly Solvable Kolezhuk-Mikeska point $X=Y=3/4$}\label{sec:ES_KM}

Recall that at $X=Y=3/4$ an exact ground state of $H(X,Y)$ in terms of a matrix product state (MPS) was found by Kolezhuk-Mikeska.\cite{Kolezhuk:prl98} The ground state at the KM point is exactly doubly degenerate even for finite system sizes. These two MPS ground states are the two possible staggered dimer coverings of the system, with no dimers between the $S$ and $\tau$ chains. Moreover, they each individually break the translational symmetry of the system. Thus, their ES may not be momentum resolved as was discussed in Section \ref{sec:symm}. In addition, these wavefunctions are exact product states of simpler spin and orbital wavefunctions. Hence they are expected to yield no entanglement whatsoever under a rung or orbital trace. These ground states are also examples of a non-Haldane gapped spin-liquid with string order.\cite{Nersesyan:prl97,Kolezhuk:prl98}

\begin{figure}
\centering
\includegraphics[width=0.5\columnwidth]{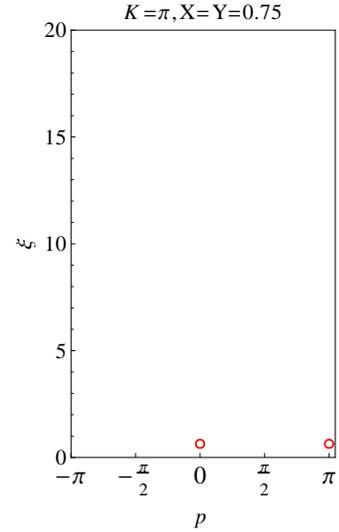}
\caption{(color online) Entanglement spectrum of an $L=12$ system from the exactly solvable Kolezhuk-Mikeska(KM) point $X=Y=3/4$ for $K=\pi$. This point is in the non-Haldane phase with a finite string order parameter.\cite{Nersesyan:prl97} There are only two levels $\xi=\ln 2$ at $p=0,\pi$. The $K=0$ spectrum (not shown) is almost identical except the levels are shifted slightly from $\ln 2$. Away from the KM point, the spectrum becomes ``dressed" with high energy levels as shown in Fig.\ref{Fig:ES_generic}(b) where there is a finite entanglement gap $\Delta\xi_\text{low}$ that separates the two sets of energies.}
\label{Fig:ES_KM}
\end{figure}

Nevertheless, by taking symmetric and antisymmetric superpositions of these exactly staggered dimers, momentum eigenstates with $K=0$ and $K=\pi$ which have the full symmetries of $H(3/4,3/4)$ are recovered. In fact, the numerical ground state wavefunctions computed around the KM point are always momentum eigenstates and are thus adiabatically connected to one of these momentum eigenstates. Interestingly, the entanglement is non-trivial for these $K=0,\pi$ eigenstates at the KM point. Shown in Fig.\ref{Fig:ES_KM} is the ES at the KM point for $K=\pi$ which is extraordinarily simple because it consists of only two levels with $\xi=\ln 2$ at momenta $p=0,\pi$ regardless of system size $L$. When $K=0$ there are still only two levels at $p=0,\pi$ but these levels are shifted slightly from $\ln 2$ by an $L$ dependent amount which vanishes in the thermodynamic limit. It thus appears that at the KM point there is an infinite entanglement gap ($\Delta\xi_\text{low}=\infty$) which becomes finite when moving away from this point as was observed in generic spectra in phase IV. 

The entanglement levels at the KM point for $K=0,\pi$ can be computed exactly using the staggered dimer wavefunctions. This calculation is outlined in the rest of this subsection. We denote a dimer between sites $i$ and $j$ on a given chain by $D_{ij}$ with the following sign convention
\begin{align}
\left| D_{ij} \right\rangle \frac{1}{\sqrt{2}}\left( |\uparrow_i\downarrow_j\rangle - |\downarrow_i\uparrow_j\rangle\right).
\end{align}
For sites $1,2,\dots,2N=L$ there is canonical dimer covering of the chain which we denote by
\begin{align}
\left|\mathbf{D}^{(1)}\right\rangle := \left| D_{12}D_{34}\dots D_{2N-1,2N}\right\rangle = \left| \prod_{i=1}^N D_{2i-1,2i}\right\rangle,
\end{align}
and its translated counterpart
\begin{align}
\left|\mathbf{D}^{(2)}\right\rangle := \left| D_{23}D_{45}\dots D_{2N,1}\right\rangle = \left| \prod_{i=1}^{N-1} D_{2i,2i+1} D_{2N,1}\right\rangle.
\end{align}
The overlap between the two distinct dimer coverings is non-zero and can be shown to be
\begin{eqnarray}
\left\langle \mathbf{D}^{(1)} | \mathbf{D}^{(2)}\right\rangle = 2(-1/2)^N.\label{eqn:Doverlap}
\end{eqnarray}
Pictorially, this says when the two coverings are overlaid on top of one another to form an unbroken chain of length $2N$, then the non-zero overlap is $2(-1/2)^N$, with our convention. This is a special case of a more general formula for general dimer coverings of a lattice. Note also that as $N\rightarrow \infty$, the overlap vanishes but is never exactly zero for finite $N$. By taking symmetric and antisymmetric superpositions and normalizing, we can form $K=0,\pi$ momentum eigenstates from the two inequivalent exact staggered dimer wavefunctions: 
\begin{align}
|\Psi_0\rangle &:= \frac{2^{N-1}}{\sqrt{2(4^{N-1}+1)}}&\nonumber \\
&\quad\quad\times \left( |\mathbf{D}^{(1)}\rangle \otimes | \mathbf{D}^{(2)}\rangle + |\mathbf{D}^{(2)}\rangle \otimes |\mathbf{D}^{(1)}\rangle\right),& \nonumber \\
|\Psi_\pi\rangle &:= \frac{2^{N-1}}{\sqrt{2(4^{N-1}-1)}}& \nonumber \\
&\quad\quad\times \left( |\mathbf{D}^{(1)}\rangle \otimes | \mathbf{D}^{(2)}\rangle - |\mathbf{D}^{(2)}\rangle \otimes |\mathbf{D}^{(1)}\rangle\right).&
\end{align}
Their respective reduced density matrices can then be computed to arrive at an $N$ dependent result for $\rho_\text{red}[\Psi_0]$ but a simple expression for $\rho_\text{red}[\Psi_\pi]$:\footnote{The reduced density matrices are expressed in a orthogonal basis obtained by Gram-Schmidt orthogonalization of $\mathbf{D}^{(1)}$ and $\mathbf{D}^{(2)}$.}

\begin{align}
&\rho_\text{red}[\Psi_0] =& \nonumber \\
&\frac{1}{2(4^{N-1}+1)}
\begin{pmatrix}
3+4^{N-1} & 2(-1)^N\sqrt{4^{N-1}-1} \nonumber \\
2(-1)^N\sqrt{4^N-1} & 4^{N-1}-1 
\end{pmatrix},& \nonumber \\ \nonumber \\
&\rho_\text{red}[\Psi_\pi] = 
\begin{pmatrix}
1/2 & 0 \\
0 & 1/2 
\end{pmatrix}.&
\end{align} 

\begin{table}[h!]
\begin{center}
    \begin{tabular}{ | l | l | l |}
    \hline
    $L=2N$ & Numerical results & Computed from $\rho_\text{red}[\Psi_0]$ \\ \hline
    8 & 0.473085, 0.975714 & 0.473085, 0.975714 \\ \hline
    12 & 0.63258, 0.757621 & 0.63258, 0.757621 \\ \hline
    16 & 0.677645, 0.708896 & 0.677644, 0.708895 \\ \hline
    \end{tabular}
   \caption{Comparison between analytical result for $\xi$ and the numerics for the $K=0$ groundstate at $X=Y=3/4$.}\label{tab}
\end{center}
\end{table}

Thus, for a $K=\pi$ ground state at $X=Y=3/4$, we expect that $1/2=\mathrm{e}^{-\xi}$ will produce two degenerate entanglement energies with value $\ln 2$. This is indeed observed in our numerics for $L=2N=8,12,16$. With the above expression for $\rho_\text{red}[\Psi_0]$, the entanglement energies can be determined and compared with our exact diagonalization results. These values are shown in Table \ref{tab} and are in almost perfect agreement.

Lastly it is clear that as $L\rightarrow \infty$, the overlap (\ref{eqn:Doverlap}) tends to zero and the differences in the ES between $|\Psi_0\rangle$ and $|\Psi_\pi\rangle$ should become negligible. 

\subsubsection{Entanglement Spectra for the SU(4) point}

\begin{figure*}
\centering
\includegraphics[width=2\columnwidth]{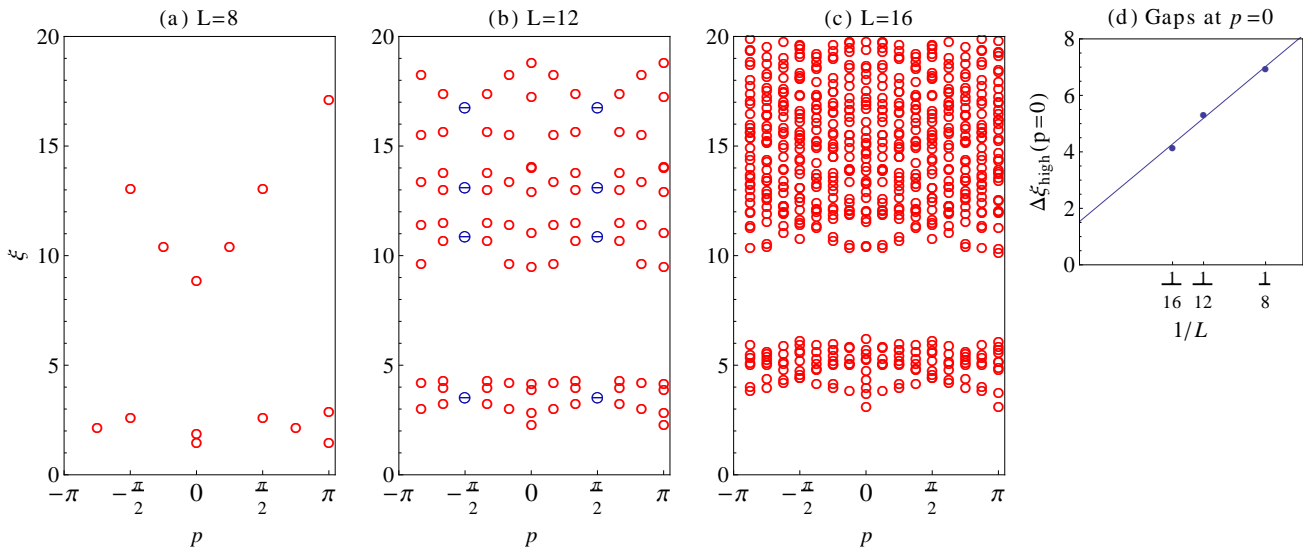}
\caption{(color online) Entanglement spectra at the $X=Y=1/4$ SU(4) point at various system sizes. The {\color{red}$\circ$} symbols denote singlet levels while the {\bf\color{DarkBlue}\scriptsize$\ominus$} symbols denote doublets. (a) $L=8$. (b) $L=12$. (c) $L=16$. The doublets at $p=\pm \pi/2$ for $L=12$ are due to the ground state momentum $K=\pi$. In all cases the total number of low lying energies below the entanglement gap obey the empirical relation $2^{L/2-1}$. (d) A finite size scaling analysis of the entanglement gap $\Delta\xi_\text{high}$ taken at $p=0$. The positive intercept suggests that the gap remains open in the thermodynamic limit.}
\label{Fig:ES_SU4}
\end{figure*}

In this subsection we focus on the rung cut ES at the SU(4) symmetric point at $X=Y=1/4$. Shown in Fig.\ref{Fig:ES_SU4}(a-c) are spectra taken from ground states at that point for sizes $L=8,12$ and $16$. A noteworthy feature shared by the spectra is the entanglement gap $\Delta\xi_\text{high}$ around $6\lesssim\xi\lesssim10$ across all momenta $p$, and the manner in which the gap becomes more well defined with increasing system size. Moreover a simple finite size scaling analysis of the entanglement gap taken at $p=0$ plotted in Fig.\ref{Fig:ES_SU4}(d) suggest that the gap remains open in the thermodynamic limit. It is possible that the entanglement gap closes logarithmically which can not be seen from our system sizes, however the overall structure of the entanglement spectra across phase boundaries (see Fig.\ref{Fig:ES_XY}) is suggestive of it remaining open.

Several features are worth pointing out regarding the systematic changes in the ES with increasing $L$. Firstly, the $L=8$ ES of Fig.\ref{Fig:ES_SU4}(a) is severely limited by the finite system size effects. Recall that $N_\text{singlet}$, the dimension of the $S^2_\text{tot}=0$ subspace of $\mathcal{H}_S$, sets the maximum number of physically relevant levels of the ES. For $L=8$, $N_\text{singlet}=14$ and this is the total number of levels observed in Fig.\ref{Fig:ES_SU4}(a). Of these $14$, $8$ levels make up the set of entanglement levels below the gap. This sparsity of the non-zero eigenvalues of $\rho_\text{red}$ highlights the limitations in attempting to draw strong conclusions about the thermodynamic limit using only $L=8$ data.

Secondly, upon increasing $L$ to $12$ and $16$ the spectrum starts to fill up with levels above and below the gap. At $L=12$, doublets are observed at $p=\pm\pi$ which can be attributed to the properties of a $K=\pi$ ground state as discussed in Appendix \ref{App:ES}. More importantly, however, this filling of levels is observed to be systematic. In all sizes considered, the total number of levels below the entanglement gap follows the empirical relation $2^{L/2-1}$. In a later section we will discuss the physical significance of the $\Delta\xi_\text{high}$ entanglement gap for correlations functions and present some conjectures as to the origin of the $2^{L/2-1}$ relation.

Lastly, as was already revealed in Fig.\ref{Fig:ES_generic}(a) and Fig.\ref{Fig:ES_Z2XY}(a,b), the $\Delta\xi_\text{high}$ entanglement gap seems to be a generic property of ground states deep in the gapless AFM phase V, where the $2^{L/2-1}$ counting is also observed at these points. Viewed in this light, the SU(4) point ES appears to be a rather generic ES of the phase V. This is rather surprising given that this point is known to be a critical point of a Kosterlitz-Thouless(KT) transition.\cite{Azaria:prl99,Itoi:prb00} In fact, at points $X=Y>1/4$ beyond the SU(4) point, the ES still yields a robust $\Delta\xi_\text{high}$ entanglement gap. This fact unfortunately has not allowed us to conclusively identify the SU(4) point as being a critical point of the KT transition using only our present amount of ES data. Nevertheless, we believe that the entanglement gap is a distinguishing property of the phase V and our interpretation of our finite system size data requires taking into account the complexities of the KT transition. In the next few sections we will present our data and analyses on the evolution of the ES along several lines in the phase diagram with a focus on the region V to IV phase transition. 

\subsubsection{Entanglement Spectra Along the $X=Y$ Line}

\begin{figure*}
\centering
\includegraphics[width=2\columnwidth]{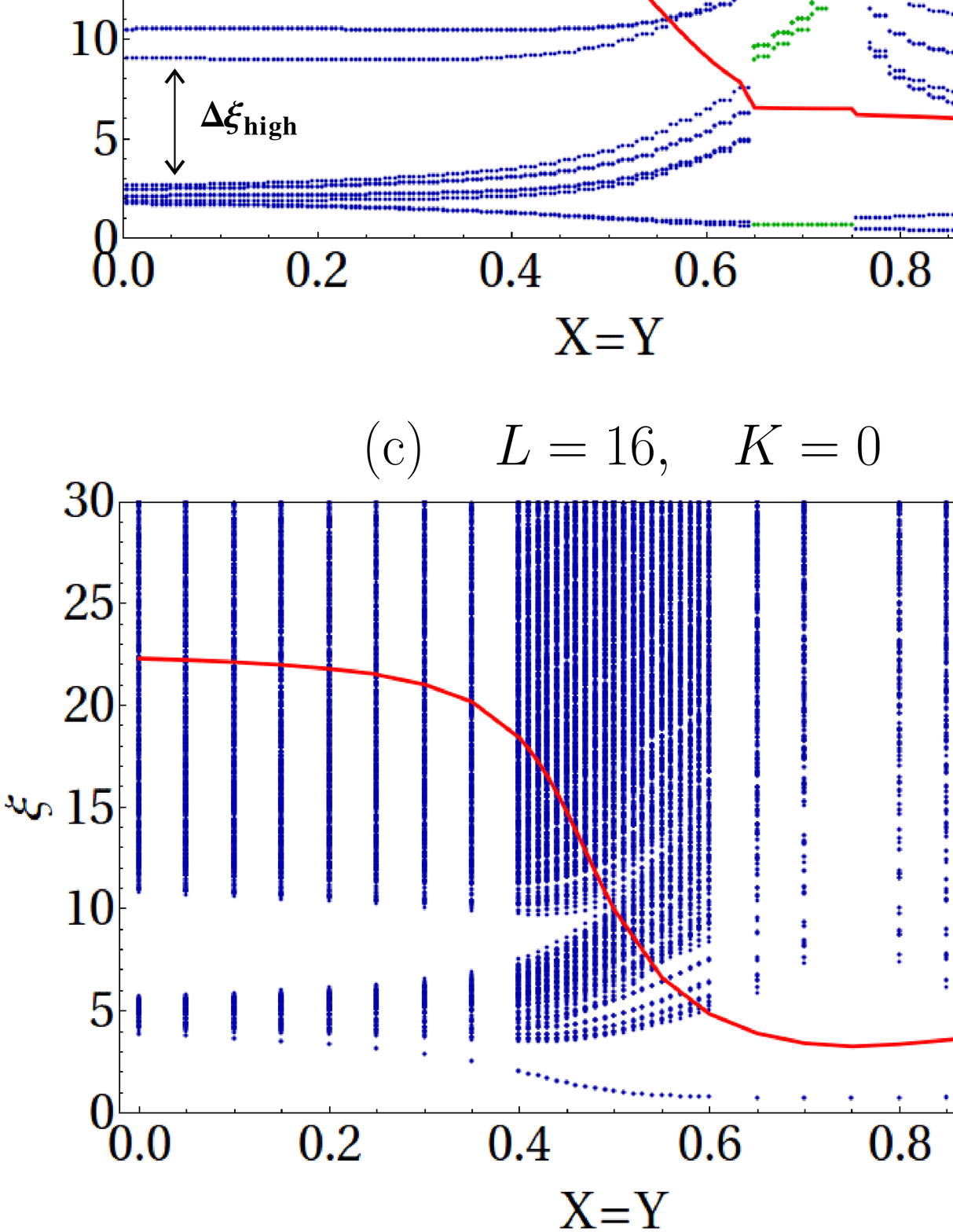}
\caption{(color online) The (non-momentum resolved) entanglement spectrum along the $X=Y$ symmetric line for sizes $L=8,12$ and $16$. The blue (dark) points correspond to spectra from ground states with momentum $K=0$ and the green (light) to $K=\pi$. In (a) and (b) the absolute numerical ground state is used to derive an ES. Due to computing limitations, our data points for $L=16$ are more sparse and we have plotted separately the ES for $K=0$ and $K=\pi$ sector ground states in (c) and (d). In phase V the absolute ground state is mostly the $K=0$ one (See Fig.\ref{Sent_E0}(b)). Overlaid as the red (solid) line is $S_\text{ent}(X,X)$ in arbitrary units. Marked out in (a) is the high entanglement gap $\Delta\xi_\text{high}$ that separates the set of $2^{L/2-1}$ low energy levels from the rest of the entanglement spectrum. Also marked in (a) is the low entanglement gap $\Delta\xi_\text{low}$ that separates the 2 distinguished lowest levels of the gapped AFM phase from the rest of the spectrum. The region where the entanglement entropy rapidly drops lies in the region where the transition from the gapless region V into the gapped regions IV and VI in Fig.\ref{pd} occurs.}
\label{Fig:ES_XY}
\end{figure*}

In this subsection we investigate the evolution of the entanglement spectrum along the $\mathbb{Z}_2$ symmetric line $0 \leq X=Y \leq 1$ for sizes $L=8,12$ and $16$. Shown in Fig.\ref{Fig:ES_XY} are the non-momentum resolved entanglement spectra along this line and overlaid with $S_\text{ent}$ data. The data for $L=8$ and $L=12$ is presented in Fig.\ref{Fig:ES_XY}(a) and Fig.\ref{Fig:ES_XY}(b). For these plots, only absolute ground states are used to compute the ES. The spectra are also labeled according to their momentum $K=0,\pi$. Our $L=16$ data set is unfortunately more limited due to computing limitations. In this case we have decided instead to show separately the spectra taken from ground states of definite momentum $K$ sectors. These results are shown in Fig.\ref{Fig:ES_XY}(c) ($K=0$) and Fig.\ref{Fig:ES_XY}(d) ($K=\pi$). As is shown in Fig.\ref{Sent_E0}(b), the $K=0$ ground state is often the absolute ground state for most of the $0 \leq X=Y \leq 1$ line.   

We remark first on the $L=8$ and $L=12$ results of  Fig.\ref{Fig:ES_XY}(a,b). In Fig.\ref{Fig:ES_XY}(a) where $L=8$ the total number of levels is limited to $N_\text{singlet}=14$ whereas in Fig.\ref{Fig:ES_XY}(b) where $L=12$ the spectrum is more densely populated. Both plots very clearly reveal the level crossings between $K=0$ and $K=\pi$ ground states. These are easily identified with discontinuous changes in the spectrum and are in exact correspondence with the positions of the non-smooth changes in the entanglement entropy $S_\text{ent}$. Otherwise, away from the level crossings, the evolution of the entanglement spectrum is continuous. This is again another demonstration of the advantage of studying the ES over the entropy when trying to detect changes in the ground state. In Figs.\ref{Fig:ES_XY}(a,b) the high entanglement gap $\Delta\xi_\text{high}$ is well defined and continuously evolves far beyond the SU(4) point $X=Y=1/4$ until a level crossing is encountered. The low entanglement gap $\Delta\xi_\text{low}$ by contrast starts to develop around $X=Y\sim 0.2$. Eventually it diverges at the exactly solvable KM point $X=Y=3/4$ where the ES is composed of just two levels. In addition, there are also parameter regions where both $\Delta\xi_\text{high}$ and $\Delta\xi_\text{low}$ are simultaneously non-zero. Lastly, we note that generally a large $\Delta\xi_\text{high}$ corresponds to a large $S_\text{ent}$.

Next, data on the ES for $L=16$ shown in Figs.\ref{Fig:ES_XY}(c,d) is significantly more sparse. Nevertheless in Fig.\ref{Fig:ES_XY}(c) for the case where $K=0$, we have have computed many more data points in the intermediate region between phase V and IV with a focus on the phase transition. The ES from this set of ground states is continuous over the entire range of $X=Y$. It is observed that the two lowest levels at $p=0,\pi$ from the KM point continuously evolve into the two lowest levels of the $2^{L/2-1}=128$ set of energies below $\Delta\xi_\text{high}$ in the phase V. In Fig.\ref{Fig:ES_XY}(c) the gap $\Delta\xi_\text{high}$ is observed to close around $X=Y\sim 0.5$ which is before the KM point but beyond the SU(4) point. At this point, the entropy $S_\text{ent}$ is also seen to decrease rapidly with increasing $X=Y$. Based on this gap closure which occurs for $L=16$ but not smaller sizes, we conjecture that with increasing $L$, the closure point of $\Delta\xi_\text{high}$ will tend towards smaller $X=Y$ values. It is known that the real energy excitation gap closes only exponentially slowly when approaching the KT critical point from phase IV.\cite{Azaria:prl99,Yasufumi:jpsj00,Itoi:prb00,Azaria:prb00} This makes it difficult to pinpoint the critical point of the transition without having to resort to a renormalization group analysis. Our numerics reflect this reality too, as determining the critical point by the opening or closure of either $\Delta\xi_\text{low}$ and/or $\Delta\xi_\text{high}$ does not lead to a satisfactory determination of the phase boundary. Perhaps future work utilizing Density Matrix Renormalization Group (DMRG) to access larger $L$ sizes with appropriate finite size scaling analyses respecting the peculiarities of the KT transition will improve on this. 
 
Finally, we should mention that $\Delta\xi_\text{high}$ is also well defined for the $K=\pi$ ground state (Fig.\ref{Fig:ES_XY}(d)) in the range $0 \leq X=Y\leq 1/4$. However these eigenstates are excited states for those parameters and interestingly, the set of low energy levels it defines actually exceeds the $2^{L/2-1}$ limit. 

Recent work has numerically substantiated the claim that one can adiabatically connect two phases as long as an entanglement gap remains open.\cite{Thomale_AC:prl10}  If that conjecture is true, one would conclude that the $\Delta\xi_\text{low}$ gap being lower in position in the spectrum and thus more significant than $\Delta\xi_\text{high}$, should open there and remain open throughout regions IV and VI. Again due to difficulties of the infinite order KT transition and the local spin-orbital degrees of freedom per site, this conjecture is hard to verify via exact diagonalization for the system sizes we could handle.  Nevertheless, the continuous and open $\Delta\xi_\text{low}$ gap (for fixed $K$) throughout region IV and VI provides further evidence that the phase diagram in Ref.[\onlinecite{Chen:prb07}] contained an unphysical phase boundary. Based on this we can again conclude with our ES results that the ground states of IV and VI are adiabatically connected once degeneracy due to dimerization is removed by fixing $K$. 

\subsubsection{Entanglement Spectra Along the Line $X+Y =0.4$}

\begin{figure}
\centering
\includegraphics[width=\columnwidth]{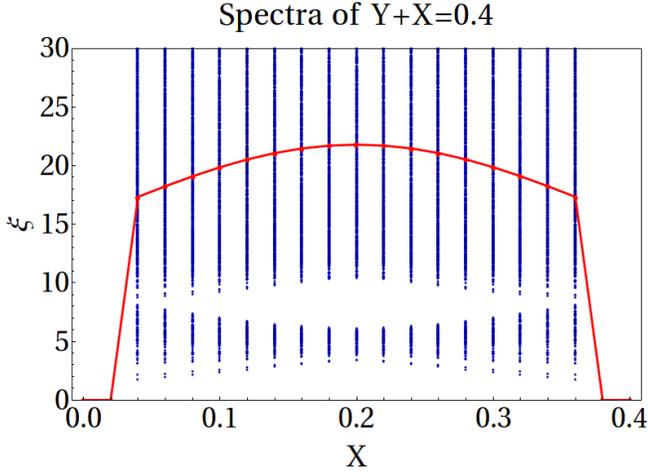}
\caption{(color online) The (non-momentum resolved) entanglement spectrum along the $X + Y = 0.4$ line for $L = 16$. The red (solid) line is the entanglement entropy. For $X \lesssim 0.06$ and $X \gtrsim 0.36$ the structure of the entanglement spectra collapses as these states have trivial entanglement. (They are the FM phases II and III of Fig.1.) For $0.06 \lesssim X \lesssim 0.36$ the entanglement gaps $\Delta\xi_\text{high}$ and $\Delta\xi_\text{low}$ are seen. This path corresponds to a slice through region V.}
\label{Fig:ES_X_Y16K0}
\end{figure}

We next consider the behavior of the (non-momentum resolved) entanglement spectra along the line $X+Y=4/10$ for $0\leq X \leq 0.4$ which is perpendicular to the symmetric $X=Y$ line. This line begins in region II of Fig.\ref{pd} and slices through through region V before finishing in region III.  Shown in Fig.\ref{Fig:ES_X_Y16K0} are the non-momentum resolved entanglement spectra along the line $X+Y=0.4$. The spectra plotted are taken from the ground state for the system size $L=16$. Results for other sizes are qualitatively similar. 

For $X\lesssim 0.06$ and $X \gtrsim 0.36$ the structure of the entanglement spectra collapses to a singular level at $\xi=0$ with $S_\text{ent}=0$ because these states have trivial entanglement.  (They are in region II and III, respectively, of Fig.\ref{pd}.) While for $0.06 \lesssim X \lesssim 0.36$, entanglement gaps $\Delta\xi_\text{high}$ and to a lesser extent $\Delta\xi_\text{low}$ are seen. This path corresponds to a slice through region V. $\Delta\xi_\text{high}$ is largest in the ``middle" of region V and appears to close at the phase boundaries. Thus, as claimed in Sec.\ref{sec:ES_generic} it appears that the $\Delta\xi_\text{high}$ gap in the entanglement spectra with the rung cut persists throughout the gapless region V.  (Recall that the SU(4) point sits on the boundary of region V in the thermodynamic limit.)

\subsubsection{Correlation Functions and the $\Delta\xi_\text{high}$ Entanglement Gap}\label{Sec:Corr}

In this subsection, we explore the physical significance of the entanglement gap $\Delta\xi_\text{high}$ for ground state(GS) correlation functions. The greatest consequence of a large well defined $\Delta\xi_\text{high}$ is the ability for certain GS correlation functions to be well approximated by using only the $N(L):=2^{L/2-1}$ Schmidt vectors below $\Delta\xi_\text{high}$. Recall that tracing over $\mathcal{H}_\tau$ to form $\rho_\text{red}$ does not remove information from observables that only depend on the spin($S$) degrees of freedom. In fact, it is these correlations that can be well approximated. 

To be more specific, let $\hat{O}(x_1,\ldots,x_m)$ be an operator depending only on $\vec{S}_{x_1},\ldots,\vec{S}_{x_m}$. Then its GS expectation value can be computed from 

\begin{equation}
\langle \Psi | \hat{O}(x_1,\ldots ,x_m) |\Psi\rangle = \text{Tr}_{\mathcal{H}_S}\{\hat{O}(x_1,\ldots ,x_m)\rho_\text{red}[\Psi] \}.
\label{eqn:corr}
\end{equation}
Let the set of entanglement levels $\{\xi_n\}$ with their corresponding set of Schmidt vectors $\{\phi_n\}$ be organized in ascending order $\xi_1\leq \xi_2 \leq \ldots \leq \xi_{N_\text{singlet}}$. Using this eigenbasis of $\rho_\text{red}$, the GS expectation can be computed with 
\begin{equation}
\langle \Psi | \hat{O}(x_1,\ldots ,x_m) |\Psi\rangle = \sum_{n=1}^{N_\text{singlet}}\mathrm{e}^{-\xi_n}\langle \phi_n | \hat{O}(x_1,\ldots,x_m)|\phi_n\rangle.
\label{eqn:corr_full}
\end{equation}
Now, a well defined gap $\Delta\xi_\text{high}$ separates the ES into low and high subsets. Moreover, the levels above $\Delta\xi_\text{high}$ will have Schmidt weights that are exponentially smaller than those below it. This can be exploited to approximate $\langle\hat{O}\rangle$ by,
\begin{equation}
\langle \Psi | \hat{O}(x_1,\ldots ,x_m) |\Psi\rangle \approx \sum_{n=1}^{2^{L/2-1}}\mathrm{e}^{-\xi_n}\langle \phi_n | \hat{O}(x_1,\ldots,x_m)|\phi_n\rangle
\label{eqn:corr_partial}
\end{equation}
with exponentially small corrections of order $O(\mathrm{e}^{-(\xi_{N(L)}+\Delta\xi_\text{high})})$. That is, with a large $\Delta\xi_\text{high}$, one only requires $N(L)=2^{L/2-1}$ Schmidt vectors to reproduce all the GS correlation functions involving only spin($S$). Although $N(L)$ still grows exponentially with $L$, it is still slower than $N_\text{singlet}$. Moreover, these GS correlation functions should obey the predictions of the SU(2)$_2$ WZW field theory that describes the low energy dynamics of the spin sector.\cite{Azaria:prl99,Itoi:prb00,Azaria:prb00} Phrased another way, the Gibbs ensemble\cite{Fiete:pra03} $\rho_\text{red}[\Psi]$ of size $N(L)$ with the degrees of freedom of a spin-1/2 chain of length $L$ can simulate well all the GS correlators of a SU(2)$_2$ WZW CFT, provided $\Delta\xi_\text{high}$ is large enough. 

\begin{figure}
\centering
\includegraphics[width=\columnwidth]{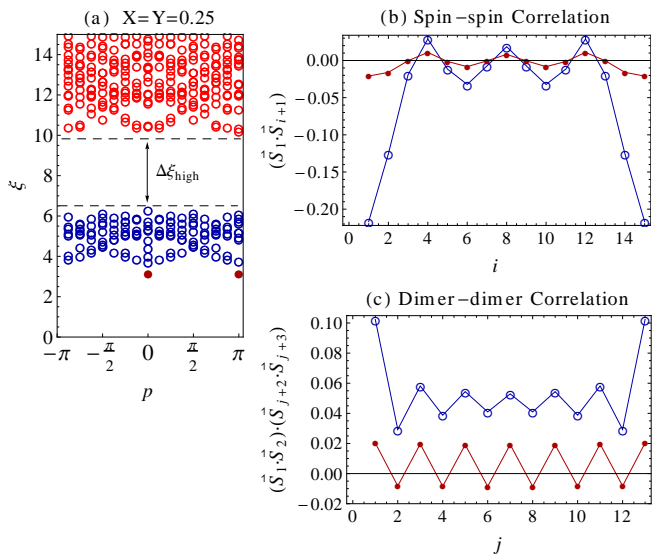}
\caption{(color online) Correlations functions taken from the SU(4) symmetric point ground state for $L=16$. (a) The entanglement spectrum where the blue {\bf\color{DarkBlue}$\circ$} symbols denote the $N(16)=128$ levels below the entanglement gap $\Delta\xi_\text{high}$. The two lowest levels with the greatest Schmidt weight are marked by the solid red {{\color{BrickRed}$\bullet$}} symbols. These levels are adiabatically connected to the two levels of the KM spectrum. (b) The spin-spin ground state correlation function. (c) The dimer-dimer ground state correlation function. In (b) and (c) the open circles {\bf\color{DarkBlue}$\circ$} are computed using only levels below $\Delta\xi_\text{high}$ while the solid {{\color{BrickRed}$\bullet$}} use only the two lowest levels of (a).}
\label{Fig:Corr}
\end{figure}

Next, we discuss our numerical results on the GS correlation functions calculated from the Schmidt vectors or equivalently the eigenvectors of $\rho_\text{red}[\Psi]$. Shown in Fig.\ref{Fig:Corr}(b,c) are spin-spin $(\vec{S}_1\cdot\vec{S}_{i+1})$ and dimer-dimer $(\vec{S}_1\cdot\vec{S}_{2})(\vec{S}_{j+2}\cdot\vec{S}_{j+3})$ ground state expectation values for a $L=16$ chain at the SU(4) symmetric point. Although we only present data for the SU(4) point, our results remain qualitatively similar at other points deep in phase V. We focus first on the open circles {\bf\color{DarkBlue}$\circ$} which are correlation functions computed using Eq.(\ref{eqn:corr_partial}). We have checked that these correlation functions are indistinguishable from those obtained using Eq.(\ref{eqn:corr_full}) where all the Schmidt vectors are utilized. Thus, we have confirmed the validity of the approximation of Eq.(\ref{eqn:corr_partial}). 

In Fig.\ref{Fig:Corr}(b) a decay is observed in the spin-spin correlation function which is unfortunately limited by (periodic) boundary condition effects. We also observed period 4 oscillations which corresponds to a $\pi/2$ peak in the structure factor. These oscillations are consistent with previous DMRG calculations\cite{Yamashita:prb98} and field theory studies.\cite{Azaria:prl99} However, due to the limited size $L=16$, our data does not allow for a fit to the $3/2$ power law decay predicted by the SU(2)$_2$ WZW field theory.\cite{Itoi:prb00,Azaria:prl99} Nevertheless, the profile is in qualitative agreement with Ref.[\onlinecite{Yamashita:prb98}] and gives us confidence that we can expect to see the algebraic decay with longer chains using only the approximation (\ref{eqn:corr_partial}). In Fig.\ref{Fig:Corr}(c) the dimer-dimer correlation function does not seem to display a strong decay like in the spin-spin case. Rather, it almost appears to be long-range ordered. But we believe this to be a finite-size effect because dimerization is not expected to occur at the SU(4) symmetry point.  

We also considered the contributions to the GS correlators from just the two lowest entanglement energy levels. That is, we calculate (\ref{eqn:corr_full}) with just the first two Schmidt vectors $\phi_1$ and $\phi_2$ which have the greatest Schmidt weight. The results of these computations as shown as the solid {{\color{BrickRed}$\bullet$}} circles in Fig.\ref{Fig:Corr}(b,c). These two levels are special in that they are smoothly connected to the entanglement levels of the KM point as shown in Fig.\ref{Fig:ES_XY}(c). Most interestingly, the resulting correlator in Fig.\ref{Fig:Corr}(c) demonstrates that their contribution dimer-dimer correlation function is surprisingly uniform and strongly long-range ordered. We have also computed the same correlation function at the KM point and a qualitatively similar profile is obtained except there the negative values at even separations is heavily suppressed. Also, the spin-spin correlator of the KM point is ultra short-ranged, decaying to zero beyond nearest neighbors. Based on these observations, we interpret the role of the remaining $N(16)-2=126$ Schmidt vectors with higher entanglement energies, as providing the necessary critical algebraic spin-spin fluctuations that destroy this long-range dimerization order. This interpretation is also based on the expectation that the SU(2)$_2$ WZW low energy effective description describes algebraic spin-liquid behavior without dimerization. This also means that $\Delta\xi_\text{low}$ is a better indicator of the gapped non-Haldane AFM phase IV, since it separates the two lowest KM levels and the next $N(L)-2$ levels. It is also worth repeating here that in certain excited states in phase V, the $\Delta\xi_\text{high}$ can sometimes be observed, but in those circumstances the number of levels below it exceeds $N(L)$. This suggests that with excited states which have more complicated correlations, there is an increased amount of complexity which manifests as an increased number of Schmidt vectors below $\Delta\xi_\text{high}$.  

We still lack any explanation, even if only intuitive of the $\Delta\xi_\text{high}$ entanglement gap and the $N(L)=2^{L/2-1}$ levels below it. This is in contrast to the $\Delta\xi_\text{low}$ gap which can been understood in terms of perturbative corrections to the KM ground state which makes the otherwise infinite gap, finite. It would be very interesting to find a ground state which is adiabatically connected to region V where $\Delta\xi_\text{high}=\infty$ such that Eq.(\ref{eqn:corr_partial}) becomes exact. We can expect such a ground state to describe the SU(2)$_2$ WZW model ground state more faithfully. Curiously, the number $N(L)$ may have a combinatorical aspect to it. We noticed the following combinatorical re-expression of $N(L)$,
\begin{equation}
N(L)=
\begin{pmatrix}
L/2\\0  
\end{pmatrix} +
\begin{pmatrix}
L/2\\2  
\end{pmatrix} +
\ldots +
{\begin{pmatrix}
L/2\\L/2  
\end{pmatrix}}. 
\end{equation}  
From this point of view, $N(L)$ is then the number of ways to select an even number of $L/2$ objects. $L/2$ is also the number of nearest neighbor links or possible dimers. We conjecture that this relation may be of some significance towards explaining the $\Delta\xi_\text{high}$ gap, although this is highly speculative at the moment. 

In most systems studied to date, an entanglement gap (with a different cut from the one considered here) has been shown to separate low-lying states with universal topological properties from higher states that contain information about excitations.\cite{Sterdyniak:11,Chandran:prb11,Cirac:prb11,Peschel:epl11}  In these cases, the systems have a gapped bulk excitation spectrum.  It is thus somewhat unusual to find an entanglement gap in a gapless system.  However, the one-dimensional spin-1/2 Heisenberg model with a cut in momentum space is an important example of a gapless system with an entanglement gap.\cite{Thomale:prl10}  In that case,\cite{Thomale:prl10} it was shown that the gap could be related to a fractional quantum Hall state of the Laughlin type, as they both share an underlying U(1) conformal field theory.

In Sec.\ref{sec:MC}, we study the entanglement spectra of the model \eqref{eq:KK} at the SU(4) point with a momentum space cut.  Quite interestingly we do \emph{not} find a gap with this cut, but we are limited by computational resources to only $L=8,12$ and there are many subtleties associated with finite-size effects, as we noted in Sec.\ref{sec:RC} A.  Nevertheless, it would be interesting to see if the entanglement spectrum from the rung cut and momentum cut could still be related to a quantum Hall system derived from  an underlying SU(4)$_1$ conformal field theory, in an analogous way to the U(1) conformal theory of the spin-1/2 Heisenberg spin chain connecting it to the Laughlin quantum Hall states.\cite{Thomale:prl10}

\subsubsection{Perturbing the Kugel-Khomskii Hamiltonian \eqref{eq:KK} with a Rung Coupling}

In a number of states with gapped bulk excitations and some type of topologically protected boundary excitations, such as fractional quantum Hall states,\cite{Li:prl08,Regnault:prl09,Zozulya:prb09,Lauchli:prl10,Thomale_AC:prl10,Sterdyniak:prl11,Thomale:prb11} topological insulators,\cite{Turner:prb10,Hughes:prb10,Kargarian:prb10,Prodan:prl10,Alexandradinata:prb11,Fiete:pe11} topological superconductors,\cite{Fidkowski:prl10,Dubail:prl11} and two-leg ladders realizing the Haldane gap phase,\cite{Pollmann:prb10,Thomale:prl10,Poilblanc:prl10,Lauchli:prb12} a real space cut of the system into ``halves" will show a momentum dependence of the entanglement eigenvalues that mimics the physical boundary excitations.  In the case of a two-leg ladder of spin-1/2 Heisenberg chains with a rung coupling,\cite{Poilblanc:prl10,Lauchli:prb12} the ``edge" would be one leg of the two.  Indeed, tracing out one leg (similar to our trace over orbital degrees of freedom in the rung cut) results in an entanglement spectrum that closely resembles that of the spin-1/2 Heisenberg model, which is gapless.\cite{Poilblanc:prl10,Lauchli:prb12}  

It is interesting to point out that at $X=Y=3/4$ (see Fig.\ref{Fig:ES_XY}(c)), a gapped phase--the non-Haldane phase with finite string order,\cite{Nersesyan:prl97,Kolezhuk:prl98} one does not see a characteristic spectra of the ``edge", which is a spin-1/2 Heisenberg chain.  This difference compared to the rung coupling is intimately related to the very different nature of the ``four-spin" interaction term $(\vec{S_i}\cdot\vec{S}_{i+1})( \vec{\tau_i}\cdot\vec{\tau}_{i+1})$ which drives rather different physics between chains.\cite{Nersesyan:prl97,Hijii:prb09}

In order to study the competition between the rung coupling and the ``four-spin" term $(\vec{S_i}\cdot\vec{S}_{i+1})( \vec{\tau_i}\cdot\vec{\tau}_{i+1})$ on the entanglement spectra in the gapped phase at the exactly solvable point $X=Y=3/4$ of Eq.\eqref{eq:KK}, we consider a perturbation of the form,\cite{Hijii:prb09}
\begin{equation}
H_\text{rung}=J_\text{rung} \sum_{i=1}^L \vec{S}_i \cdot \vec{\tau}_i.
\end{equation}
In the spin-ladder interpretation of $\mathcal{H}_{S\tau}$, this term is akin to an AFM exchange coupling between rungs of the ladder and hence the reason we have chosen to refer to it as rung coupling. This term breaks the global SU(2)$\times$ SU(2) rotation symmetry down to just SU(2). The angular momentum symmetry properties of the ground state(GS) are thus modified too. It is no longer necessarily true that $S^2_\text{tot}=\tau^2_\text{tot}=0$ but only that $(\vec{S}_\text{tot}+\vec{\tau}_\text{tot})^2=0$. This then leads to the entanglement spectrum having support in more than one $S^2_\text{tot}$ sector of $\mathcal{H}_S$ and allows the possibility of higher spin angular momentum multiplets in the entanglement spectrum. 

\begin{figure}
\centering
\includegraphics[width=\columnwidth]{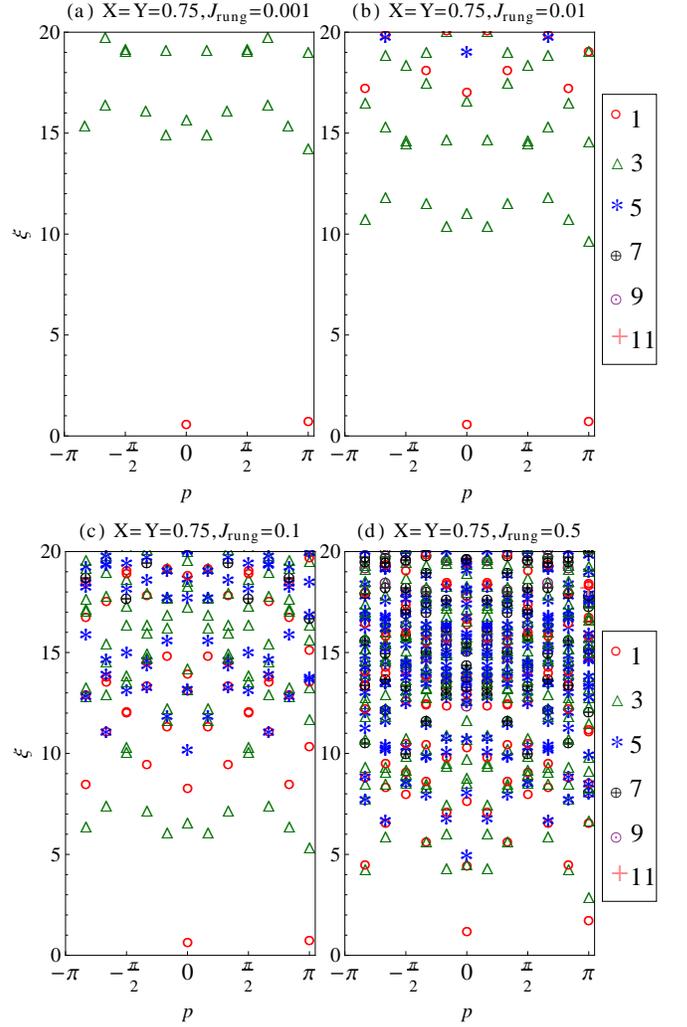}
\caption{(color online) Momentum resolved entanglement spectra for $L=12$ chains at the KM point with $J_\text{rung}$ coupling (a) $J_\text{rung}=0.001$. (b)$J_\text{rung}=0.01$. (c) $J_\text{rung}=0.1$. (d) $J_\text{rung}=0.5$. In (d) the spectrum resembles the energy spectrum of a Heisenberg spin-1/2 chain and the ES of a two-leg ladder without the four spin or dimer-dimer interaction term. The degeneracies of the levels are denoted by the symbols displayed in the legend.}
\label{Fig:ES_KM_J}
\end{figure}

First, we consider the effect of $H_\text{rung}$ on the entanglement spectrum of the KM point where our results are shown in Fig.\ref{Fig:ES_KM_J}. In Fig.\ref{Fig:ES_KM_J}(a), we observed that even with a very small $J_\text{rung}>0$ relative to $X$ and $Y$, the ES exhibits triplets and higher odd multiplets that descend from infinite entanglement energy, thus rendering $\Delta\xi_\text{low}$ finite. This ES is significantly different from the ES obtained by just perturbing $X$ or  $Y$ from the KM point, where only singlet or doublets are observed due to symmetries. However, both share a $\Delta\xi_\text{low}$ gap which indicates that they are in the same phase. It should also be noted that when $J_\text{rung}=0.001$ as in Fig.\ref{Fig:ES_KM_J}(a), the ground state is still expected to be in the same gapped topological non-Haldane phase. This is not as surprising as it seems since even with a pure spin-1/2 two leg Heisenberg ladder without the four spin-interaction term, changes in the multiplet structure of the excited entanglement energies can occur without going through a phase transition.\cite{Poilblanc:prl10,Lauchli:prb12} In Fig.3 and Fig.4 of Ref.[\onlinecite{Poilblanc:prl10}], entanglement spectra were computed for the rung singlet (I) and rung singlet (II) phases which exhibit different lowest level multiplets (triplets and sextets respectively). Yet, the ground state between these ``phases" is a gapped product states of rung dimers and is not strictly at a phase transition. Hence based on this and our results, we conclude that  comparing the multiplet structure of lowest level ES excitations of a rung cut is not a well suited method to distinguish between gapped topological quantum phases.  

Returning to Fig.\ref{Fig:ES_KM_J}, with increasing $J_\text{rung}$ the gap $\Delta\xi_\text{low}$ diminishes and more complicated degenerate entanglement levels appear in the spectrum. The largest rung coupling value that we considered $J_\text{rung}=0.5$ shown in Fig.\ref{Fig:ES_KM_J}(d) somewhat resembles the degeneracy structure of the ES of two leg Heisenberg ladder in the Haldane and rung singlet(I) phase\cite{Poilblanc:prl10} which mimics the physical spectrum of the spin-1/2 Heisenberg model.\cite{Poilblanc:prl10,Lauchli:prb12} But differences appear in the details of the position of the levels and the overall dispersion. Nevertheless, in the large $J_\text{rung}>0$ limit, we expect to recover a ground state similar to the rung singlet(I) phase and we should recover the results of Ref.[\onlinecite{Poilblanc:prl10}].

\begin{figure}
\centering
\includegraphics[width=\columnwidth]{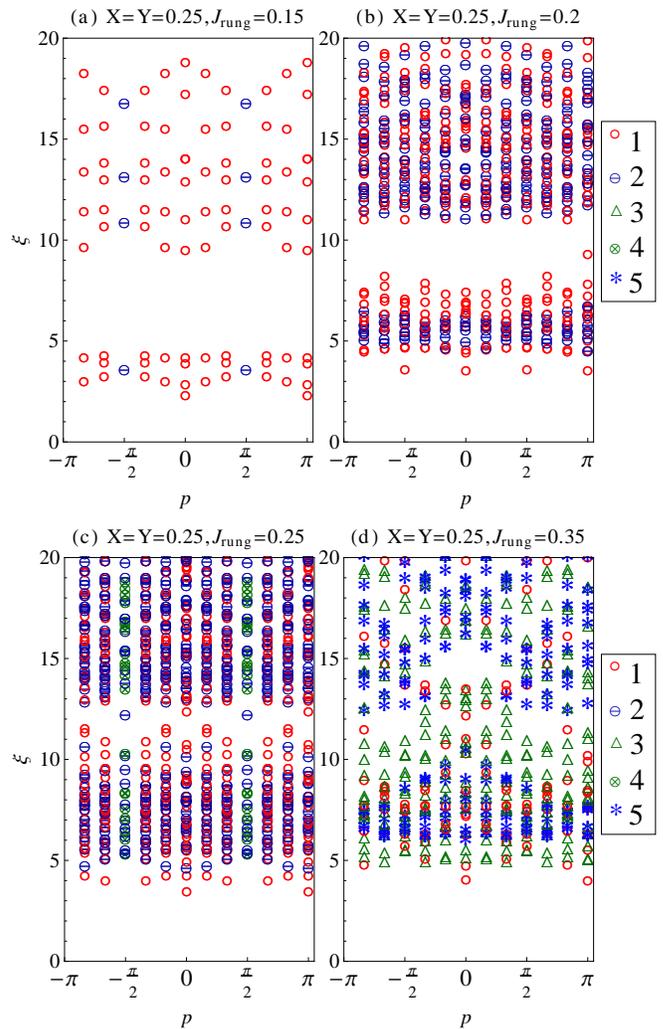}
\caption{(color online) Momentum resolved entanglement spectra for $L=12$ chains at the SU(4) point with $J_\text{rung}$ coupling (a) $J_\text{rung}=0.15$. (b)$J_\text{rung}=0.2$. (c) $J_\text{rung}=0.25$. (d) $J_\text{rung}=0.35$.  The degeneracies of the levels are denoted by the symbols displayed in the legend.}
\label{Fig:ES_SU4_J}
\end{figure}

Finally, we consider perturbing around the SU(4) symmetric point and our results are presented in Fig.\ref{Fig:ES_SU4_J}. Remarkably the ES is surprisingly robust and remains roughly unchanged from the SU(4) point ES even when $J_\text{rung}=0.15$ as shown in Fig.\ref{Fig:ES_SU4_J}(a). Nevertheless with increasing $J_\text{rung}$ dramatic changes start to appear, indicative of some sort of phase transition or crossover. In Fig.\ref{Fig:ES_SU4_J}(b) with $J_\text{rung}=0.2$, there is a proliferation of doublets which occur, even at momentum away from $p=\pm\pi/2$. There, $\Delta\xi_\text{high}$ is still somewhat open. Increasing $J_\text{rung}$ further causes the $\Delta\xi_\text{high}$ to collapse and eventually the even multiplets give way to only odd multiplets in the ES. We observed that the final spectrum in Fig.\ref{Fig:ES_SU4_J}(d) does not vary much with further increases in $J_\text{rung}$ and the low entanglement energy dispersion appears to be rather flat.

These results on perturbations about the exactly solvable KM point and the SU(4) symmetric point are still poorly understood. But they do highlight the rich structure that is revealed by the entanglement spectrum when changes to the ground state occur, especially in the case of perturbing the KM ground state. 

\section{Bond or Leg Cut Entanglement Spectrum}
\label{sec:BC}
In order to obtain complementary information on the entanglement properties of the one-dimensional Kugel-Khomskii Hamiltonian \eqref{eq:KK}, we now consider a cut in the Hilbert space where the system is cut in ``half" lengthwise\cite{Eisert:rmp10,Amico:rmp08} through two bonds in real space of the periodic chain. This partition then defines two contiguous blocks of sites each with length $L/2$, where one of the blocks is traced over. In the spin-ladder interpretation, this partition cuts the legs of the ladder symmetrically. 

This is the most common partition used for gapped systems, and is natural for revealing the real boundary excitations of an open chain in the Haldane phase, for example.\cite{Pollmann:prb10} We study system sizes of $L=8$ and $12$. In a symmetry protected topological phase, like the Haldane phase, the lowest levels of the entanglement spectrum will be two-fold degenerate per virtual edge due to free spin-1/2 edge states.\cite{Pollmann:prb10} This even degeneracy is protected by various symmetries, including inversion symmetry.\cite{Liu:prb11,Pollmann:prb10} Note that with the periodic boundary conditions we use there are always two virtual edges. The quantity of interest is then the parity of the degeneracies per edge as it characterizes the localized edge excitations of the entanglement Hamiltonian $H_\text{ent}$.\cite{Turner:prb11,Raussendorf:np10}

\begin{figure}
\centering
\includegraphics[width=0.5\columnwidth]{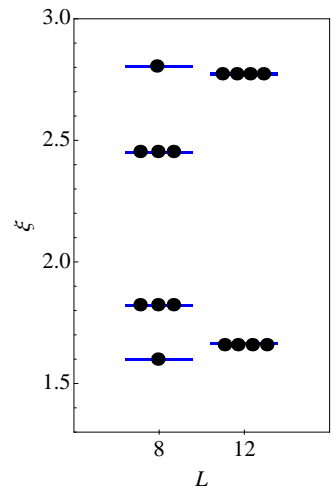}
\caption{(color online) The entanglement spectrum at the KM point plotted versus system size. The black dots mark the degeneracy of each level. For 8 sites, there is a residual effective coupling between virtual edge excitations in $H_\text{ent}$ that splits the four fold degenerate states into singlets and triplets. For 12 sites where the edges are more spatially separated the 4-fold degenerate or 2-fold degenerate per edge is restored to a good approximation. The eigenvalues are equal up to the third decimal place.}
\label{KMPoint_cutinhalf}
\end{figure}

We first investigate the entanglement spectrum at the KM point. Picking a single MPS groundstate out of the two degenerate ones that spontaneously breaks the translational symmetry,\cite{Kolezhuk:prl98} we used the analytical formalism developed in Ref.[\onlinecite{Pollmann:prb10}]. We verified that the entanglement spectrum is four-fold degenerate (two per virtual boundary) in the thermodynamic limit.\footnote{One must double the unit cell to use the analytical techniques in Ref.[\onlinecite{Pollmann:prb10} ]} One can also intuitively see this from considering the wavefunction which is a staggered dimer covering of a two leg ladder. Then a bond or leg cut will always cut a singlet dimer (which could be a spin or orbital dimer) at each edge which leads to the two fold degenerate free spin-1/2 levels on each edge. Thus the KM point entanglement spectrum only has one set of two-fold degenerate eigenvalues per virtual edge, exactly as the Affleck-Kennedy-Lieb-Tasaki\cite{Affleck:prl87} state. However, the arguments developed in Ref.[\onlinecite{Pollmann:prb10}] break down for degenerate ground states, which is the case for the KM point. Nevertheless, this even degeneracy when properly interpreted is still symmetry protected and should remain throughout the non-Haldane phase. As was emphasized earlier, generically in a finite size system the dimerization is never exact and one obtains almost degenerate ground states with definite linear momentum $K=0,\pi$. In this case we find that in the neighborhood of the KM point, there are \emph{two} sets of nearly four-fold degenerate entanglement levels as opposed to one. These numerical results are presented in Fig.\ref{KMPoint_cutinhalf}. But this may be accounted for by taking linear combinations of the MPS states like was done in Section \ref{sec:ES_KM} to arrive at definite $K$ ground states. Then one can see through linearly superposing the individual Schmidt decompositions of the exact MPS states that the final reduced density matrix now should have two sets of four-fold degenerate entanglement levels which is still even per virtual edge. By adiabatic continuity, if one limits to fixed $K$ ground states these degeneracies should remain invariant as long as the real excitation gap remains finite. We emphasize that this added complication in the entanglement spectrum at finite system size is really another artifact of the dimerization order exhibited by the non-Haldane phase and a proper interpretation requires taking into account this degeneracy by choosing a definite linear momentum $K$. 

This special degeneracy can be used to identify the topological order in this state, but since it is the same as in the Haldane phase the bond cut alone does not allow one to distinguish the Haldane phase from the non-Haldane phase. The failure of the degeneracy of the entanglement spectrum to identify different topological phases has been pointed out before.\cite{Liu:prb11}  (It is possible that a ``particle partition" \cite{Sterdyniak:11} entanglement spectrum might reveal the difference between the Haldane and non-Haldane phase as their excitations are different.\cite{Nersesyan:prl97}) However, our experience with the rung cut leads us to believe that a combination of a leg and a rung cut could provide a way to distinguish many different topological gapped phases of two leg ladder systems or spin chains with orbital degeneracy. In order to distinguish between the non-Haldane phase and the Haldane phase one could first search for a degeneracy to establish the presence of a topological phase, and then study the rung cut entanglement spectrum to see if there are two low-lying levels and an entanglement gap (non-Haldane) or the energy spectrum of a single Heisenberg spin chain (Haldane). Combining and comparing different partitions is not limited to spin ladder systems.  It can be applied to other systems to gain complimenatry information.  However, it is beyond the scope of this paper (and may not even be possible) to obtain a systematic approach to determine the minimum number and specific type of partitions needed to fully classify a given system.
 
We also study the bond cut entanglement spectrum at the SU(4) point. At this point the continuum limit of the entanglement spectrum is completely determined\cite{Calabrese:pra08} by the central charge, $c=3$. For a finite gapless system, the mean number of eigenvalues of the reduced density matrix larger than a given eigenvalue, $\lambda$ , is given by 
\begin{equation}
\label{eq:n}
n(\lambda)=I_0(2\sqrt{b \ln(\lambda_{max}/\lambda)}),
\end{equation}
where $I_0$ is the modified Bessel function, $b=\frac{c}6  \ln(\frac{L}{a}\sin(\pi \frac{l}{L}))$, $c$ is the central charge, and $l$ is the subchain length. For simplicity we let $z=2\sqrt{b \ln(\lambda_{max}/\lambda)})$.

\begin{figure}
\centering
\includegraphics[width=\columnwidth]{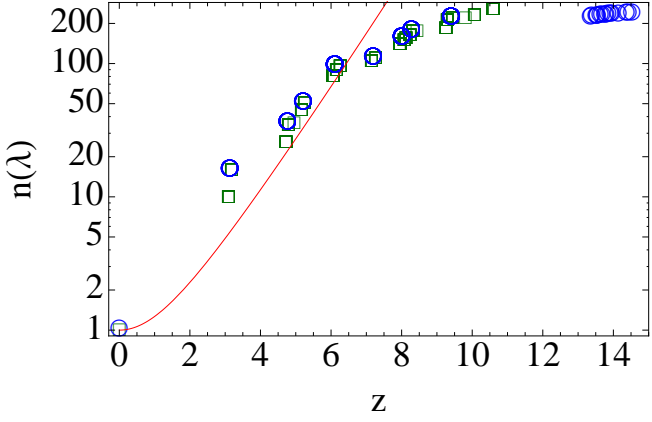}
\caption{(color online) The mean number of eigenvalues, $n(\lambda)$, of the reduced density matrix larger than a given eigenvalue, $\lambda$, plotted as a function of $z=2\sqrt{b \ln(\lambda_{max}/\lambda)})$ in Eq.\eqref{eq:n} for $l=4$ and $L=12$. The blue circles represent data for $X=0.25$. The lowest level at $z=0$ is non-degenerate and the next level is 15-fold degenerate. The green squares represent data for $X=0.2$. The lowest level at $z=0$ is non-degenerate and the next level is 9-fold degenerate. The red line is the analytical prediction.\cite{Calabrese:pra08}}
\label{cuthalf_25}
\end{figure}

In Fig.\ref{cuthalf_25} we investigate how the low energy excitations of \eqref{eq:KK} effects the degeneracy of the entanglement spectrum in the gapless region V in Fig.\ref{pd}. We first compute the entanglement spectrum at the SU(4) point. This is done a chain of length 12 by tracing over 8 sites. A subchain of length 4 is chosen to better reflect the thermodynamic limit.  After the lowest entanglement level we observe an entanglement level with a 15-fold degeneracy. If we break the SU(4) symmetry by moving along the $X=Y$ line to $X=0.2$, the symmetry is reduced to $\mathrm{SU}(2)\ifmmode\times\else\texttimes\fi{}\mathrm{SU}(2)\ifmmode$. We observe the 15-fold degeneracy splits into 9-fold and 6-fold degenerate sets of levels. The entanglement levels that are 9-fold degenerate have a lower entanglement energy and are thus more entangled. We can relate these degeneracies of the entanglement spectrum to the degeneracies of the lowest energy excitations. At the SU(4) point, excitations with quantum numbers $(S^2,\tau^2)=(0,1),(1,0),(1,1)$ are degenerate\cite{Yasufumi:jpsj00} (where ``1" means triplet) and transform under the fundamental irreducible representation of SU(4) and hence are 15-fold degenerate. At $X=Y=0.2$, however the lowest energy excitation is the $(S^2,\tau^2)=(1,1)$ mode which transform as two independent triplets under SU(2)$\times$ SU(2). This then gives the 9-fold degeneracy. At that point the excitations with quantum numbers $(S^2,\tau^2)=(1,0),(0,1)$ are the next lowest excitations and they together are degenerate and give the 6-fold degenerate excitations. This is somewhat surprising since the ground state is in the gapless phase. Unfortunately, our system sizes are too small for our data to fall nicely on the curve \eqref{eq:n} derived for the continuum limit. However, both the $n(\lambda)$ and the dependence of the entanglement entropy (not shown) on sub-system length are consistent with $c=3$ given the size limitations.

\section{Momentum Cut}
\label{sec:MC}

In our final section of the paper, we consider a cut defined with momentum space ``orbitals" of the Hilbert space (which is very non-local in position space) at the SU(4) point.  This momentum-space cut has earlier been used to reveal interesting structure in the entanglement spectrum, including a clear entanglement gap and CFT counting of levels in the spin-1/2 Heisenberg chain system that reflect the gapless bulk excitations,\cite{Thomale:prl10} and has also been used in characterizing disordered fermion systems. \cite{taylor_momentum_cut} In this section, we extend this momentum space partition of Ref.[\onlinecite{Thomale:prl10}] to a spin chain with orbital degeneracy or equivalently a two leg spin ladder.

To define this cut, we first write the ground state wavefunction in a Holstein-Primakoff representation,
\begin{align}
&|\Psi\rangle=\nonumber \\
&\sum_{\substack{j_1 < ... < j_\kappa \\ i_1 < ... < i_\kappa}}\Psi (z_{j_1},..., z_{j_\kappa}|w_{i_1},..., w_{i_\kappa}) S^-_{j_1}... S^-_{j_\kappa}\tau^-_{i_1}... \tau^-_{i_\kappa}|F\rangle,\quad\quad\label{eqn:HolsteinPrimakoff}
\end{align}
where $|F\rangle = |\uparrow \ldots \uparrow\rangle$ is the fully polarized state and $\kappa=L/2$ is number of spin ($S$) and pseudospin ($\tau$) flip operators. $\kappa$ is set by the polarization condition $S^z_\text{tot}=\tau^z_\text{tot}=0$ that is satisfied by the ground state. We have also expressed the real space wavefunction coefficients in U(1) coordinates of the periodic chain with $z_j:=\mathrm{e}^{i\frac{2\pi j}{L}}$ and $w_j:=\mathrm{e}^{i\frac{2\pi j}{L}}$ being positions of a lowered spin and a pseudospin respectively. Note that the complex valued function $\Psi(z_{j_1},...|w_{i_1},...)$ is totally symmetric in all its arguments and is zero whenever any two arguments are equal. This property derives from the mutual commutativity of these spin flip operators and that their square is always zero, $(S_j^-)^2=(\tau_i^-)^2=0$. Next, we define the Fourier and inverse Fourier transforms of these spin flip operators by 
\begin{eqnarray}
\tilde{S}^-_m:=\frac{1}{\sqrt{L}}\sum_{j=1}^L \bar{z}_j^m S^-_j, \quad \quad 
S^-_j=\frac{1}{\sqrt{L}}\sum_{m=1}^L z^m_j \tilde{S}^-_m, \nonumber \\
\tilde{\tau}^-_{\check{m}}:=\frac{1}{\sqrt{L}}\sum_{i=1}^L \bar{w}_i^{\check{m}} \tau^-_i, \quad\quad
\tau^-_i=\frac{1}{\sqrt{L}}\sum_{\check{m}=1}^L w^{\check{m}}_i \tilde{\tau}^-_{\check{m}}.\nonumber \\
\label{eqn:FT}\end{eqnarray}
The ground state wavefunction may then be expressed in terms of these Fourier transformed operators as
\begin{align}
&|\Psi\rangle =\nonumber\\
&\sum_{\substack{m_1,...,m_\kappa\\ \check{m}_1,...,\check{m}_\kappa}}\tilde{\Psi}(m_1,...,m_\kappa|\check{m}_1,...,\check{m}_\kappa)
\tilde{S}^-_{m_1}...\tilde{S}^-_{m_\kappa}\tilde{\tau}^-_{\check{m}_1}...\tilde{\tau}^-_{\check{m}_\kappa}|F\rangle,
\label{eqn:FT1}\end{align}
where the sum over the different ``momenta" $m_j$ and $\check{m}_i$ takes values in ${\{0,1,\ldots,L-1\}}$ but is otherwise unrestricted. In the rest of this section we will measure the crystal momentum in units of $2\pi/L$. The coefficients of (\ref{eqn:HolsteinPrimakoff}) and (\ref{eqn:FT1}) are related to each other according to
\begin{align}
&\tilde{\Psi}(m_1,...,m_\kappa|\check{m}_1,...,\check{m}_\kappa) = \nonumber \\
&\frac{1}{L^{\kappa}}\sum_{\substack{j_1<...<j_\kappa \\ i_1<...<i_\kappa}}\Psi(z_{j_1},...,z_{j_\kappa}|w_{i_1},...,w_{i_\kappa})z^{m_1}_{j_1}...z^{m_\kappa}_{j_\kappa}w^{\check{m}_1}_{i_1}...w^{\check{m}_\kappa}_{i_\kappa}. \label{eqn:FT2}\end{align} 
Note that $\tilde{\Psi}$ remains symmetric under the action of the permutation group of $\kappa$ objects, $S_\kappa$ on $(m_1,...,m_\kappa)$ and $(\check{m}_1,...,\check{m}_\kappa)$. However, the spin flip operators $\tilde{S}^-_m$ and $\tilde{\tau}^-_{\check{m}}$ in Eq.(\ref{eqn:FT}) may be freely permuted because they are all mutually commuting and no longer square to zero unlike their real space counterparts. Thus in the sum over $m$'s and $\check{m}$'s, we should only consider terms that are unique. This leads to a description of states which resembles an occupation number basis for bosonic particles. Thus inspired by the usual bosonic Fock space, we define a set of basis states by
\begin{eqnarray}
|\mathbf{n},\mathbf{\check{n}}\rangle&:=& \prod_{m=0}^{L-1} \frac{(\tilde{S}^-_m)^{n_m}(\tilde{\tau}^-_m)^{\check{n}_m}} {\sqrt{n_m!\check{n}_m!} }|F\rangle,
\end{eqnarray}
where $\mathbf{n}:=(n_0,\ldots,n_{L-1})$ and $\mathbf{\check{n}}:=(\check{n}_0,\ldots,\check{n}_{L-1})$. The spin flip operators of (\ref{eqn:FT1}) may be thought of as ``magnon"-like creation operators acting on the ``vacuum" $|F\rangle$ and $(\mathbf{n},\mathbf{\check{n}})$ corresponds to an occupation configuration of such magnon orbitals.\cite{Thomale:prl10} However $[\tilde{S}^-_m,\tilde{S}^+_m] \propto \tilde{S}^z_m$ and similarly for $\tilde{\tau}_{\check{m}}^\pm$. Hence, these are not strictly speaking bosonic creation operators. Moreover, the basis defined by such states is non-orthogonal and overcomplete, although they are spin, pseudospin and momentum eigenstates. Nevertheless, we shall refer to the orbital states created by $\tilde{S}^-_m$ and $\tilde{\tau}^-_{\check{m}}$ as magnons and regard the occupation basis as orthonormal. With this in mind, the ground state wavefunction may be uniquely written in the occupation basis of magnons as follows
\begin{equation}
|\Psi\rangle = \sum_{\mathbf{n},\mathbf{\check{n}}} \tilde{\Psi}(\mathbf{n},\mathbf{\check{n}}) |\mathbf{n},\mathbf{\check{n}}\rangle,
\end{equation}
with
\begin{align}
&\tilde{\Psi}(\mathbf{n},\mathbf{\check{n}}) =\prod_{m=0}^{L-1}\frac{1}{\sqrt{n_m!\check{n}_m!}}\tilde{\Psi}(m_1,...,m_\kappa|\check{m}_1,...,\check{m}_\kappa),
\end{align}
where $m_1\leq m_2 \ldots \leq m_\kappa$ is a magnon configuration representative of the occupation $\{n_0,...,n_{L-1}\}$ and similarly for $\check{m}_i\leq...\leq\check{m}_\kappa$ and $\{\check{n}_m,...,\check{n}_{L-1}\}$. However, these occupation numbers are not entirely unrestricted as symmetries of the wavefunction will impose constraints on their space of possibilities. For example, the $S^z_\text{tot}=\tau^z_\text{tot}=0$ condition constrains their sum such that
\begin{equation}
\sum_{m=0}^{L-1}n_m=\sum_{m=0}^{L-1}\check{n}_m=\kappa.
\label{eqn:sum_rule}
\end{equation}
Due to time-reversal symmetry, the ground state may only have momentum $K=0$ or $K=L/2=\kappa$ in units of $2\pi/L$. This then imposes a constraint on the total momentum
\begin{align}
&M_\text{tot}:=M_A+M_B=\sum_{m=0}^{L-1}m\times (n_m +\check{n}_m), \nonumber 
\end{align} 
such that
\begin{align}
&M_\text{tot} \;\text{mod}\; L= K, 
\end{align}
where $K$ is measured in units of $2\pi/L$. Here, $M_A:=\sum_{m=0}^{L/2-1}m(n_m+\check{n}_m)$ and $M_B:=\sum_{m=L/2}^{L-1}m(n_i+\check{n}_m)$ are the total right and left moving momenta from both spin($S$) and orbital($\tau$) magnons. Finally, we introduce the bipartite partition in momentum orbital space by splitting the orbitals into right movers with $m,\check{m}=0,...,L/2-1$ (partition $A$) and left movers with $m,\check{m}=L/2,...,L-1$ (partition $B$). The reduced density is then formed by tracing out the orbitals in $B$ in the pure ground state density matrix, that is $\rho_\text{red}=\text{Tr}_B|\Psi\rangle \langle\Psi|$. 
\begin{figure*}
\centering
\includegraphics[width=1.5\columnwidth]{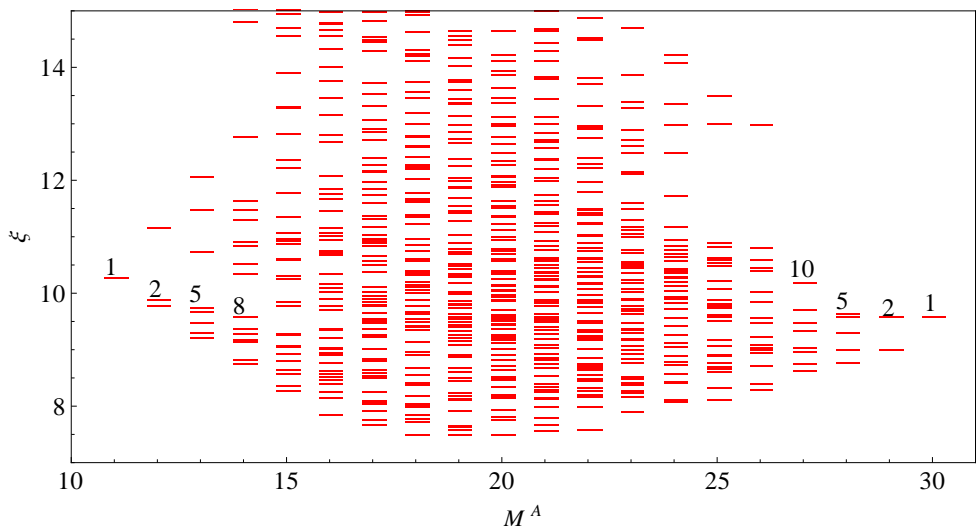}
\caption{The momentum cut entanglement spectrum at the SU(4) symmetric point for $L=12$ in the sector $(N_A,\check{N}_A)=(3,3)$. The ground state possesses momentum $K=\kappa=L/2$ in units of $2\pi/L$. The count of the lowest levels from high $M_A$ (right) is 1,2,5,10 and from low $M_A$ (left) is 1,2,5,8. The floating numbers in the plot denote the total number of levels directly below. There are two pairs of levels which are almost degenerate at $M_A=27$ which are not resolved at this scale. The spectrum does not exhibit an entanglement gap despite being in the gapless phase V described by a sum of two SU(2)$_2$ WZW theories or a single SU(4)$_1$ WZW theory.}\label{Fig:L12_ES_Magnon_SU4}
\end{figure*}

Next, we discuss the various ways in which we organized the the spectrum of $H_\text{ent}=-\ln \rho_\text{red}$ which is the entanglement spectrum. Let $N_A:=\sum_{m=0}^{L/2-1}n_m$ be the total number of spin magnons in $A$ and $\check{N}_A:=\sum_{m=0}^{L/2-1}\check{n}_m$ the total number of pseudospin magnons in $A$ and similarly for the $B$ partition. Then the sum rule of Eq.\ref{eqn:sum_rule}) implies $N_A+N_B=\check{N}_A+\check{N}_B=\kappa$, which is satisfied by each ket in the occupation basis expansion. Because of this sum rule on the total magnon numbers, $\rho_\text{red}$ will be block diagonal in $(N_A,\check{N}_A)$ blocks. In our study we focus on blocks where the $A$ partition contains half the total number of magnons, that is $N_A=\check{N}_A=\kappa/2$. Thus partition $A$ will posses a spin and pseudospin polarization of $(\Delta S^z_\text{tot},\Delta\tau^z_\text{tot})=(-N_A,-\check{N}_A)$ relative to $|F\rangle$. In addition, we block or resolve the spectrum of $\rho_\text{red}|_{(N_A,\check{N}_A)}$ into $M_A$ quantum numbers. It must be mentioned, however, that $M_A$ is not a strictly a good quantum number as was in the case of the spin-1/2 Heisenberg chain.\cite{Thomale:prl10} Only when the wavefunction weights $|\tilde{\Psi}(n_0,...)|^2$ have support in one $M_\text{tot}$ sector, can $M_A$ be a good quantum number because $M_\text{tot}$ too is a good quantum number of $\tilde{\Psi}$. Our numerical results show a distribution of weights in more than one $M_\text{tot}$ sector. Nevertheless, in regards to the lowest set of levels in the entanglement spectrum, $M_A$ remains an approximately good quantum number to label these levels. In Appendix \ref{App:MA}, we present additional numerical results to substantiate this claim. Hence, each $M_A$ sector contains a net spin and pseudospin current of $(J_S^z,J_\tau^z)=(-N_A M_A,-\check{N}_A,M_A)$ in units of $2\pi/L$. 

We present our numerical computations of the entanglement spectrum at the SU(4) point. Numerical limitations allow us to only consider systems of sizes $L=8$ and $L=12$. Shown in Fig.\ref{Fig:L12_ES_Magnon_SU4} is the momentum cut entanglement spectrum for a system of size $L=12$ at the SU(4) point in the sector $(N_A,\check{N}_A)=(3,3)$. In contrast to the a single spin-1/2 Heisenberg chain, an entanglement gap is \emph{not} exhibited in our spectrum. This goes against the expectation that the presence of gapless bulk excitations is manifested by an entanglement gap which separates a set of universal low lying entanglement energies with CFT counting of levels, as was demonstrated in the spin-1/2 Heisenberg chain.\cite{Thomale:prl10} Hence, this particular partition, as we have performed it, is ill suited to infer the existence of gapless bulk excitations from the entanglement spectrum in this model. The complications with the several $M_\text{tot}$ sectors may have been a precursor of this fact. Nevertheless, we next consider the counting of low energy entanglement levels when they are well-defined. Due to the periodic nature of momentum space (first Brillouin zone), there are really two cut regions between $m=L-1,0$ and between $m=L/2-1,L/2$. Thus we should be expect to see CFT counting of levels from either low or high momentum. However, we observe instead a counting of lowest energy levels which is asymmetric from high and low $M_A$, which again is in contrast with the single Heisenberg chain. From large $M_A$ we see a count of lowest energy levels of 1,2,5,10. From small $M_A$ where there appears to be a small gap opening we observe a count of 1,2,5,8 instead, but this may be due to finite size effects as it was not observed previously with $L=8$ data (See Fig.\ref{Fig:L8_ES_Magnon_SU4}). Moreover, the lowest energy levels of the spectrum also appears to exhibit a strong dispersion which is be bowl-like in contrast to the relatively flat set of lowest levels seen in the spin-1/2 Heisenberg chain. We note that the 1,2,5,10 sequence may be obtained from the dimensions of highest weight and descendant states of a tensor product of two independent Verma modules of the Virasoro algebra, each with the usual counting 1,1,2,3 with increasing level. Presently, we lack satisfactory interpretations of the absence of the entanglement gap and the observed counting of states in terms of the CFT that describes the low energy bulk excitations.
 
\section{Conclusions}
\label{sec:conclusions}
In conclusion, we have investigated the entanglement properties of the one-dimensional Kugel-Khomskii model \eqref{eq:KK} using exact diagonalization for system sizes of $L=8,12$ and 16 sites. By considering various partitions or ``cuts" of the Hilbert space, including a rung (orbital), bond, and momentum cut we have produced a comprehensive picture of the phase diagram from the point-of-view of the quantum entanglement. 

Of the three partitions that we have studied, the partition into spin and orbital degrees of freedom was the most intensely studied. With that partition, we have shown that there are crucial finite-size effects at $L=8$ in the entanglement entropy and the entanglement spectrum that impacted earlier exact diagonalization studies of this model which led to a misinterpretation of the phase diagram.\cite{Chen:prb07} In particular, using the entanglement spectrum, we are able to reaffirm that regions IV and VI are adiabatically connected to one another once ground state degeneracies due to dimerization are taken into account. We have observed and explained various symmetries of the entanglement spectrum obtained by this cut which can depend crucially on the ground state linear momentum $K$. More importantly we observed that the entanglement spectrum of the rung cut does not in any way resemble the energy spectrum of a single spin-1/2 Heisenberg chain which is the system one obtains when the orbital and spin degrees of freedom are decoupled. Thus the Kugel-Khomskii chain offers a counter example to the commonly held boundary-bulk correspondence conjecture of the entanglement spectrum.\cite{Chandran:prb11,Qi:prl12,Cirac:prb11,Peschel:epl11} This is intimately related to the natural $c=3$ CFT of the gapless phases, while two spin-1/2 Heisenberg edges only produce a net central charge of $c=2$. With this cut, we have also observed two important entanglement gaps. The lower gap which we have denoted as $\Delta\xi_\text{low}$ characterizes the dimerization of phase IV, which is the non-Haldane phase. $\Delta\xi_\text{low}$ separates the two lowest entanglement levels from a continuum and diverges at the exactly solvable Kolezhuk-Mikeska (KM) point.  We have computed the entanglement spectrum exactly at this special point. The second and higher gap which we have denoted by $\Delta\xi_\text{high}$ characterizes the gapless AFM phase or phase V. It separates a set of $N(L)=2^{L/2-1}$ levels from the continuum and we show through studies of the correlation functions that these levels are the source of the critical fluctuations of the ground state spin-spin correlators. Our study of the evolution of the entanglement spectra through the Kosterlitz-Thouless phase transition between phase V and IV(VI) sharply identifies the level crossings between the approximately degenerate ground states. More importantly, once $K$ is fixed as was done in Fig.\ref{Fig:ES_XY}(c), the evolution of the spectrum reveals the opening and closures of the entanglement gaps which is indicative of a phase transition. However, our finite size data is still unable to definitively identify the SU(4) point as the exact location of the critical point of the KT transition. By perturbing Eq.(\ref{eq:KK}) with a direct spin-orbital exchange perturbation, we showed that the SU(4) point rung cut entanglement spectrum and the $\Delta\xi_\text{high}$ gap is surprisingly robust. By contrast, perturbations at the KM point show a finite $\Delta\xi_\text{low}$ gap developing but reveal odd multiplets in the entanglement levels above $\Delta\xi_\text{low}$ even when remaining in the non-Haldane phase. Thus we showed that the multiplet structure of the excited entanglement levels are a poor identifier of a topological phase with this cut. 

With the bond or leg cut entanglement spectrum study, we observed in the non-Haldane phase two 2-fold degenerate entanglement levels per virtual edge which we attribute to nearly degenerate ground states due to spontaneous dimerization. In the gapless phase, we see a distribution of eigenvalues which because of small system sizes does not fit accurately to CFT predictions\cite{Calabrese:pra08} but is still consistent. Furthermore, the lowest entanglement level degeneracies reveal the degeneracies of lowest real energy excitations.   

Finally, we extended the non-local momentum cut of Thomale et. al.\cite{Thomale:prl10} to two leg ladders or spin chains with orbital degeneracy. We obtained the entanglement spectrum at the SU(4) symmetry point for a chain of size $L=12$. In contrast to the single spin-1/2 Heisenberg chain we do not see a clear entanglement gap which would be an indicator of gapless CFT bulk excitations. However we observed a new counting of levels that has yet to be identified with the SU(4)$_1$ or two SU(2)$_2$ WZW theories that are the two alternate low energy descriptions of the system.  

\acknowledgements
We thank M. Kargarian, D. Lorshbough, H. Yao, X.-L. Qi, T. Hughes, E. Ardonne, N. Regnault, A. Bernevig, and R. Thomale for useful discussions. We gratefully acknowledge financial support through ARO Grant W911NF-09-1-0527, and NSF Grant
DMR-0955778. G.A.F acknowledges the hospitality of the Aspen Center for Physics under NSF Grant PHY-1066293 where part of this work was done. R.L. was partially supported by an NSF GRF.  The authors acknowledge the Texas Advanced Computing Center (TACC) at The University of Texas at Austin for providing computing resources that have contributed to the research results reported within this paper. URL: http://www.tacc.utexas.edu

\appendix

\section{Features of the Entanglement Spectrum when $K=\pi$}\label{App:ES}

Our data on the entanglement spectrum reveals non-trivial relations between spectra and possible degeneracies when the ground state $|\Psi\rangle$ has momentum $K=\pi$. In this appendix we present arguments to explain the origin of these features.

First consider a non-degenerate ground state with generic $X \neq Y$ with $K=0,\pi$ and Schmidt decomposition,
\begin{align}
|\Psi\rangle = \sum_p\sum_{\phi_p}\mathrm{e}^{-\xi_{\phi_p}/2}|p\phi_p\rangle \otimes |(K-p)\overline{\phi_p}\rangle, 
\end{align}
and energy $H(X,Y)|\Psi\rangle=E_0|\Psi\rangle$. Then because $H(Y,X)\mathcal{F}|\Psi\rangle = \mathcal{F}H(X,Y)|\Psi\rangle=E_0\mathcal{F}|\Psi\rangle$, thus $\mathcal{F}|\Psi\rangle$ is a ground state of $H(Y,X)$ which is also non-degenerate. But 
\begin{eqnarray}
\mathcal{F}|\Psi\rangle &&=\sum_p\sum_{\phi_p}\mathrm{e}^{-\xi_{\phi_p}/2}|(K-p)\overline{\phi_p}\rangle\otimes|p\phi_p\rangle \nonumber \\
&&=\sum_p\sum_{\phi_{(K-p)}}\mathrm{e}^{-\xi_{\phi_{(K-p)}}/2}|p\overline{\phi_{(K-p)}}\rangle \otimes |(K-p)\phi_p\rangle, \nonumber \\
\end{eqnarray}
where in the last line we have relabeled the dummy momentum $p\rightarrow K-p$. Thus $\mathcal{F}|\Psi\rangle$ yields an entanglement spectrum of $\{(p,\xi_{\phi_{(K-p)}})\}$. This implies that the (non-degenerate) ground states related by $\mathcal{F}$ have their entanglement spectra shifted according to $K-p$ relative to each other. Since $K=0$ or $\pi$, this means that $|\Psi\rangle$ with spectrum $\{(p,\xi_{\phi_p})\}$ is related to the spectrum of $\mathcal{F}|\,\Psi\rangle$ with spectrum $\{(p,\xi_{\phi_{-p}})\}$ for $K=0$ or $\{(p,\xi_{\phi_{\pi-p}})\}$ for $K=\pi$. Since we have already established the $\{(p,\xi_{\phi_p})\}\equiv \{(p,\xi_{\phi_{-p}})\}$ symmetry due to inversion $\mathcal{I}$, the $K=0$ case requires that the spectra of $\mathcal{F}\,|\Psi\rangle$ and $|\Psi\rangle$ be identical. But for $K=\pi$ the shift is non-trivial and is confirmed by our numerics as shown in Fig.\ref{Fig:ES_Z2XY}(a-b).

Next focusing on $K=\pi$ and the special momenta where $\pi-p=p$ mod $2\pi$. This occurs when $p=\pm\pi/2$ and at these special momenta, the $|\Psi\rangle$ and $\mathcal{F}|\Psi\rangle$ levels are identical. 
\begin{eqnarray}
\begin{matrix}
|\Psi\rangle & &\mathcal{F}|\Psi\rangle \\
\{(\frac{\pi}{2},\xi_{\phi_{\pi/2}})\} & \equiv & \{(\frac{\pi}{2},\xi_{\phi_{\pi/2}})\} \\ 
 \|| & & \|| \\
\{(-\frac{\pi}{2},\xi_{\phi_{-\pi/2}})\} & \equiv & \{(-\frac{\pi}{2},\xi_{\phi_{-\pi/2}})\} \\ 
 \end{matrix}
\end{eqnarray}

Finally, we consider the special limit when $X=Y$. In this case $H(X,X)$ does commute with $\mathcal{F}$ and non-degenerate ground states transform into themselves modulo a sign under $\mathcal{F}$. Considering still only the case when $K=\pi$ and so we have an additional symmetry in the entanglement spectrum, 
\begin{align}
\{(p,\xi_{\phi_p})\} &\equiv \{(\pi-p,\xi_{\phi_p})\} \equiv \{(\pi+p,\xi_{\phi_{-p}})\}& \nonumber \\
&\equiv \{(\pi+p,\xi_{\phi_p})\}.&
\end{align}
Hence, $p$ and $p+\pi$ have identical spectra too and in particular $p=0$ and $p=\pi$ are identical. When $p=\pm\pi/2$ we have the empty statement that
\begin{align}
\left\lbrace(\pm\pi/2,\xi_{\phi_{\pm\pi/2}})\right\rbrace \equiv \left\lbrace(\pm\pi/2,\xi_{\phi_{\pm\pi/2}})\right\rbrace.
\end{align}
For definiteness, take $p=\pi/2$ (an identical consider applies for $p=-\pi/2$) and looking at the contribution to $|\Psi\rangle$ in the Schmidt decomposition, 
\begin{align}
\sum_{\phi_{\pi/2}}\mathrm{e}^{-\xi_{\phi_{\pi/2}}}\left|\pi/2\;\;\phi_{\pi/2}\right\rangle \otimes \left|\pi/2\;\;\overline{\phi_{\pi/2}}\right\rangle. 
\end{align}
Applying $\mathcal{F}$ yields
\begin{align}
\sum_{\phi_{\pi/2}}\mathrm{e}^{-\xi_{\phi_{\pi/2}}}\left|\pi/2\;\;\overline{\phi_{\pi/2}}\right\rangle \otimes \left|\pi/2\;\;\phi_{\pi/2}\right\rangle
\end{align}
which is nothing but the statement that spectrum at $p=\pi/2$ maps back into itself. However if we suppose $\mathcal{F}|\Psi\rangle=-|\Psi\rangle$ then we cannot have $\overline{\phi_{\pi/2}}\propto\phi_{\pi/2}$ else we will arrive at a contradiction (this however is consistent with $\mathcal{F}|\Psi\rangle=|\Psi\rangle$). Hence we must have that $\overline{\phi_{\pi/2}}\bot\phi_{\pi/2}$ and that the eigenvalue $\xi_{\phi_{\pi/2}}$ must be degenerate with $\xi_{\overline{\phi_{\pi/2}}}$. Thus, the entanglement energies must be at least two-fold degenerate at $p=\pi/2$ in this case. This then explains the doublets seen in Fig.\ref{Fig:ES_Z2XY}(d).

\section{$M_A$ as an approximately good quantum number in the momentum cut}\label{App:MA}

In this appendix we show that blocking $\rho_\text{red}|_{(N_A,\check{N}_A)}$ into total right moving momentum $M_A$ sectors has little quantitative influence on lowest levels of the full entanglement spectrum of $\rho_{red}|_{(N_A,\check{N}_A)}$. In the case of a single spin-1/2 Heisenberg chain, the expression of ground state wavefunction in terms of the magnon occupation basis yielded coefficients with weights $|\tilde{\Psi}(n_0,...)|^2$ that were sharply concentrated around sector $M_\text{tot}=\kappa^2=L^2/4$.\cite{Thomale:prl10} This led to a sum rule $M_A+M_B=\kappa$ which in turn led to $\rho_\text{red}$ being block diagonal in $M_A$. Unfortunately, in the case of the spin-orbital chain, things are more complicated and the weights are distributed over three to four sectors depending on $K$. When $K=0$ the weights are distributed over $M_\text{tot}=2\kappa^2,2\kappa^2\pm L$ and when $K=L/2$ they are distributed over $M_\text{tot}=2\kappa^2\pm \kappa, 2\kappa^2\pm \kappa \pm L$. Shown in Fig.\ref{Fig:L8_SU4_Weights} is the weight distribution of the coefficients for an $L=8$ Kugel-Khomskii spin chain ground state at the SU(4) symmetric point $X=Y=1/4$ where most of the weight is concentrated at $2\kappa^2$. 

\begin{figure}[h]
\centering
\includegraphics[width=0.8\columnwidth]{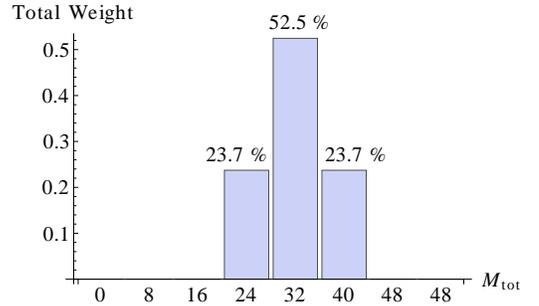}
\caption{The total weight of the occupation basis coefficients taken from the ground state at the SU(4) point. The system size is $L=8$ and the ground state momentum is $K=0$. The distribution of weights in distributed across three sectors, $M_\text{tot}=2\kappa^2,2\kappa^2\pm L$, where $\kappa=L/2=4$.}\label{Fig:L8_SU4_Weights}
\end{figure}

Nevertheless, it appears that $M_\text{tot}$ is an almost a good quantum number for $|\Psi\rangle$, at least with regards to characterizing the entanglement spectrum of $\rho_\text{red}=\text{Tr}_B |\Psi\rangle\langle\Psi|$. If we compare the spectrum obtained by just diagonalizing $\rho_\text{red}$ and the spectrum obtained by first explicitly blocking into $M_A$ sectors (ignoring off diagonal terms between sectors) and then computing the resulting spectrum, we find spectra which agree very well with one another. Shown in Fig.\ref{Fig:L8_ES_Magnon_SU4_NoBlock} are the two spectra computed with these two different methods. The data shows an agreement between the entanglement energies up till the very highest levels around $\xi\gtrsim 40$, where we believe the spectrum is less likely be of strong physical significance.  We also present in Fig.\ref{Fig:L8_ES_Magnon_SU4_NoBlock} the entanglement spectrum resolved in $M_A$ sectors for an $L=8$ chain. 
\begin{figure}
\centering
\includegraphics[width=\columnwidth]{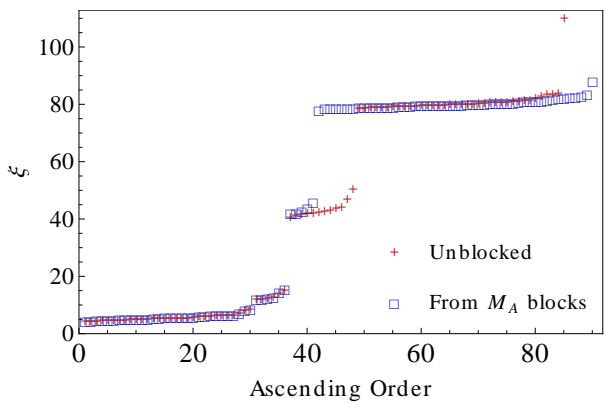}
\caption{Entanglement spectra arranged in ascending order taken from the SU(4) symmetric point of an $L=8$ chain in the sector $(N_A,\check{N}_A)=(2,2)$. Shown by the boxes ({\color{blue}$\square$}) is the spectrum from first blocking $\rho_\text{red}$ and the crosses ({\color{red}$+$}) the spectrum without blocking $\rho_\text{red}$. The two spectra agree at the lowest energies and differ at the highest levels $\xi\gtrsim 40$.}\label{Fig:L8_ES_Magnon_SU4_NoBlock}
\end{figure} 

\begin{figure}
\centering
\includegraphics[width=0.8\columnwidth]{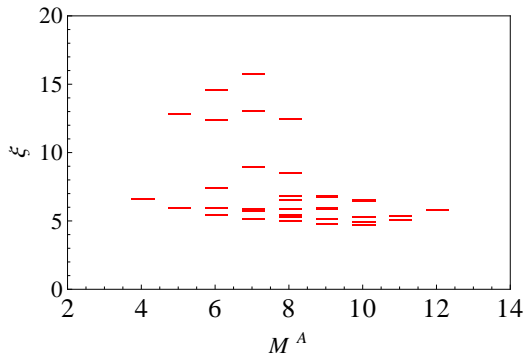}
\caption{The momentum cut entanglement spectrum of the Kugel-Khomskii chain at the SU(4) symmetric point for $L=8$. The spectrum is taken from into $M_A$ blocks of $\rho_\text{red}$ in the sector $(N_A,\check{N}_A)=(2,2)$. The count of the lowest levels from high $M_A$ (right) is 1,2,5 and from low $M_A$ (left) it is 1,1,3.}\label{Fig:L8_ES_Magnon_SU4}
\end{figure}

With a system of size $L=12$, the agreement between blocked and unblocked spectra is not as strong but the lowest entanglement energy levels nevertheless are independent of the $M_A$ blocking with significant deviations only starting to appear around $\xi \sim 12$. Moreover, we expect that only these robust lowest entanglement levels will reflect the low real energy properties of the system. 

%

\end{document}